%% file: 00main_arxiv.tex
\documentclass[nonacm,sigplan]{acmart}

\usepackage{xspace}
\newcommand{\ignore}[1]{}
\newcommand{\DESIGN}{\mbox{GOLDYLOC}\xspace}

\usepackage{algorithm}
\usepackage{algpseudocode}

\usepackage{soul}
\usepackage{letltxmacro}
\LetLtxMacro{\oldhl}{\hl}
\renewcommand{\hl}[1]{#1}                      

\usepackage{breqn}
\usepackage{amsmath}

\usepackage{soul}
\usepackage{letltxmacro}
\LetLtxMacro{\oldst}{\st}                   
\makeatletter
\renewcommand\st[1]{\@bsphack\@esphack}
\makeatother

\usepackage{tikz}
\newcommand*\circled[1]{\tikz[baseline=(char.base)]{
            \node[shape=circle,draw,inner sep=0.2pt, minimum size=0.1cm] (char) {#1};}}

\definecolor{ForestGreen}{RGB}{34,139,34}

\usepackage{subfig}
\usepackage{graphicx}

\usepackage{multirow}
\usepackage{colortbl}

\begin{document}

\title{Global Optimizations \& Lightweight Dynamic Logic for Concurrency}

\author{Suchita Pati\textsuperscript{1}, Shaizeen Aga\textsuperscript{1}, Nuwan Jayasena\textsuperscript{1} and Matthew D. Sinclair\textsuperscript{1,2}}
\affiliation{
\institution{\textsuperscript{1}Advanced Micro Devices Inc.,  \textsuperscript{2}University of Wisconsin-Madison} \country{USA}}
\email{{suchita.pati, shaizeen.aga, nuwan.jayasena}@amd.com,  {sinclair}@cs.wisc.edu}

\input{01_new_abstract}

\maketitle 
\pagestyle{plain} 

\input{02_new_intro}

\input{03_background_motivation}
\input{04_challenges}
\input{05_proposal}
\input{06_meth}
\input{07_evaluation}

\input{08_discussion}
\input{09_related}
\input{10_conc}

\bibliographystyle{plain}
\bibliography{references}

\end{document}

%% file: 01_new_abstract.tex
\begin{abstract}
Modern accelerators like GPUs are increasingly executing independent operations concurrently to improve the device's compute utilization.
However, effectively harnessing it on GPUs for important 
primitives such as general matrix multiplications (GEMMs) remains challenging.
Although modern GPUs have significant hardware and software support for GEMMs, their kernel implementations and optimizations typically assume each kernel executes in \textit{isolation} and can utilize all GPU resources.
This approach is highly efficient when kernels execute in isolation, but causes significant resource contention and slowdowns when kernels execute concurrently.
Moreover, current approaches often only \textit{statically} expose and control parallelism within an application, without considering runtime information such as varying input size and concurrent applications -- often exacerbating contention. 
These issues limit performance benefits from concurrently executing independent operations.
Accordingly, we propose \DESIGN{}, which
considers the \textit{global} resources across all concurrent operations to identify performant GEMM kernels, which we call globally optimized (GO)-Kernels.
Moreover, \DESIGN{} introduces a lightweight dynamic logic which considers the \textit{dynamic} execution environment for available parallelism and input sizes to execute performant combinations of concurrent GEMMs on the GPU.
Overall, \DESIGN{} improves performance of concurrent GEMMs on a real GPU by up to 2$\times$ (18\% geomean per workload) and provides up to 2.5$\times$ (43\% geomean per workload) speedups over sequential execution.
\end{abstract}

%% file: 02_new_intro.tex
\section{Introduction}
\label{sec:intro}

GPUs have emerged as the accelerator of choice for many domains, including machine learning (ML), as they offer a strong combination of programmability, performance, and energy efficiency.
Accordingly, GPU vendors have designed highly tuned software~\cite{cublas,rocblas,cudnn,khan2019miopen} and hardware support (e.g., Matrix Core Engines~\cite{amd-cdna-arch} and TensorCores~\cite{volta}) that accelerate common operations such as GEMMs.
As a result, GPU floating point operations per second (FLOPS) have scaled significantly across generations (e.g., 4$\times$ from 2022 to 2023~\cite{mi210,mi300x}).
Although application resource (memory, compute) requirements have also scaled~\cite{JouppiYoon2021-tpuv4, NaffzigerBeck2021-amdChiplets, MLGrowth, ShoeybiPatwary2019-megatronlm}, their individual operations often do not have high device utilization (Section~\ref{subsec:back-util}).
This is especially true for deep neural networks (DNNs) 
on GPUs.
For example, GEMMs, which make up 30-65\% of the runtime in recurrent neural networks (RNNs) and Transformer networks~\cite{VaswaniShazeer17-attention}, only utilize 40-50\% of a GPU~\cite{IvanovDryden2021-transformersDataMove, PatiAga2022-demystifying, ShoeybiPatwary2019-megatronlm, ZadehPoulos2019-dlTime}.
This occurs due to their inherent model structure (e.g., sequential processing in RNNs), low input batching to meet latency requirements, and/or due to the use of data/model partitioning techniques~\cite{jia2018data} (e.g., tensor slicing) to increase the overall memory available to the application.
As a result, significant device resources are idle in current systems for these algorithms.

\begin{table}[tb!]
  \centering
  \resizebox{1\columnwidth}{!}{%
  \input{tables/conc_capability}
  }
  \caption{Mechanisms to exploit concurrency on GPUs, including operators optimized in isolation vs. for global resources and static/dynamic concurrency management.}
  \label{tab:capability_overview}
\end{table}

One useful technique to improve compute utilization is to concurrently execute independent operations.
Programmers expose independent operations via streams~\cite{amd-hip, nvidia-stream, nvidia-stream2} within applications and use multi-instance deployments~\cite{mig, mxgpu}.
Systems typically greedily maximize the number of concurrent operations to execute.
However, naively executing independent operations concurrently can be sub-optimal and may degrade performance.
Two key factors impact this.
First, kernels must be aware of and optimized for environment they are executed in (\textit{Operator Optimization Environment}), including resources shared during concurrent executions. 
Second, operators whose performance degrades 
from sharing resources must avoid concurrent executions (\textit{Concurrency Control Logic}).
We use these factors as axes in Table~\ref{tab:capability_overview} to describe how current GPUs and prior work leverage concurrency (related work discussed further in Section~\ref{sec:relWork}).

\textbf{Current GPUs} optimize operator implementations for \textit{isolated} environments.
GPU libraries exhaustively tune implementations of key operators like GEMMs for performance (Section~\ref{subsubsec:back-gemms-impl}).
However, this tuning assumes GEMMs execute in isolation and can use all GPU resources.
It does not consider how resources may be shared \textit{globally} during execution due to potential intra- and inter-process concurrency.
Thus, while these operators are fast and efficient when executed in isolation,
when executed concurrently with other operators, resource sharing and contention can cause them significant slowdowns.
For example, concurrently running two GEMMs 
from a wide range of DNNs provides only a 10\% geomean performance improvement over sequentially executing them and 
only 7\% geomean for 16 concurrent GEMMs (detailed in Section~\ref{sec:challs}).
While resource partitioning techniques~\cite{mxgpu,OtternessAnderson2020-cuMask,mig} provide partial but dedicated resources to each concurrent operation, their benefits are limited by kernel implementations tuned for all resources.

Furthermore, current GPUs \textit{statically} manage concurrency within an application (e.g., using streams), while the hardware concurrently schedules as many operations (kernels) as possible.
However, the concurrency benefits and/or opportunities available within a device can change \textit{dynamically} with varying application inputs~\cite{PatiAga20-seqPoints} and multiple simultaneous processes.
Thus, the number of concurrent GPU kernels can be higher or lower than desired, exacerbating contention and hurting performance.
For example, concurrently running sixteen Transformer layer GEMMs with BERT~\cite{DevlinChang18-bert} model sizes improves performance by 20\%, but those with GPT-3~\cite{BrownMann2020-gpt3} sizes suffer 10\% performance degradation.
In Section~\ref{sec:challs} we show this issue also occurs in many other DNNs.

Recent work on GPU \textbf{wavefront} (WF)~\cite{cruise, Rogers2012, cawa,dwf, gpu-trace-schedule, gputhread, RogersOConnor2013-daws,stall-aware,simd_sched_gra, owl, micro_2_lvl,pats,LiuYang2015-saws} and \textbf{queue}~\cite{ChenYang2017-prophet, ChenYang2016-baymax, GaoYu2018-batchmaker, HolmesMawhirter2019-grnn,gpusync, KatoLakshmanan2011-timeGraph, YehSinclair2021-lax, adriaens2012case} \textbf{schedulers} improve upon current GPUs by \textit{dynamically} managing intra- and/or inter-process concurrency with heuristics (e.g., deadlines, synchronization, cache contention, or stalls).
However, since they use kernels optimized for \textit{isolation}, despite the number of concurrent operations, they lose out on performance benefits from implementations optimized for \textit{global} shared resources during concurrency.

Other GPU research such as \textbf{Elastic Kernels (EK)}~\cite{pai2013improving} and \textbf{Rammer}~\cite{MaXie2020-rammer} 
partially consider the \textit{global} resource environment.
EK dynamically adjusts kernels' WorkGroup/grid sizes to maximize overlap but does not apply to kernels that use shared memory or local data share (LDS)~\cite{pai2013improving} -- which GEMMs heavily utilize (Section~\ref{subsubsec:back-gemms-impl}).
Rammer re-compiles applications and their kernels to exploit operational parallelism within an application.
However, Rammer uses simple GEMM implementations unlike those in state-of-the-art BLAS libraries~\cite{rocblas,cublas}.
Furthermore, neither EK nor Rammer \textit{dynamically} manage a device's concurrency, degrading throughput in some cases. For example, Rammer can only be applied \textit{statically} to intra-application concurrency and is cumbersome for dynamic input sizes.

Collectively, these state-of-the-art schemes use a range of solutions to exploit parallelism.
However, none of them select kernels optimized for \textit{global} resource environments \textbf{and} consider \textit{dynamic} information on available parallelism, both of which are necessary to realize concurrency benefits.
Unfortunately, both globally optimized kernel implementations and dynamically controlling concurrency are challenging to realize.
GEMMs can be bottlenecked by different resources (e.g., memory, compute) during concurrency based on their input.
Furthermore, and similar to baseline BLAS libraries, each GEMM of a given size requires unique kernel implementations to optimize for the bottlenecked resource.
Manually identifying such implementations for a range of GEMMs can be challenging.
Furthermore, determining the appropriate amount and combination of concurrent operations based on available parallelism requires profiling, which can incur significant overheads when done at runtime to capture dynamic information. 
Alternately, using simple heuristics to determine the appropriate concurrency is insufficient; in Section~\ref{sec:challs} we show that a combination of multiple factors including tensor sizes, input sizes, shapes, memory layouts, and kernel implementations dictate whether and how much concurrency is beneficial.
Thus, concurrency benefits cannot be determined at runtime using simple heuristics.

Accordingly, we propose \textbf{\DESIGN{}}.
\DESIGN{} augments kernel tuning to identify, for each input, efficient kernels for both isolation and \textit{global} shared resource environments resulting from varying degrees of concurrent execution.
To find the latter, \DESIGN{} tunes kernels offline with \textit{resource constraints}, which emulates various shared resource environments.
Similar to the baseline, isolated-tuned, BLAS libraries where kernels have unique properties per GEMM input, tuning for concurrency also makes unique trade-offs per input to efficiently share resources while limiting a GEMM's performance degradation.
To \textit{dynamically} select the appropriate kernels at runtime based on the global resource environment and concurrency, \DESIGN{} extends the kernel scheduling data structure to include pointers to globally optimized kernels.
This allows the GPU's command processor (CP), the interface between software and hardware, 
to select the appropriate kernel at runtime.
Moreover, we augment the GPU's CP to dynamically control the executed concurrency using a predictor (trained offline) to select the appropriate concurrency to exploit -- i.e., which type and degree of concurrent GEMMs to select given the available independent GEMMs and their inputs.
This includes detecting if sequential execution is preferred when concurrency hurts performance.
To our knowledge, \DESIGN{} is the first to combine \textit{dynamic} concurrency control and \textit{globally} optimized GPU kernels.

We evaluate \DESIGN{} on a real GPU using the open-source BLAS infrastructure from AMD~\cite{Tensile,rocblas}.
Overall, across 410 GEMMs from modern DNNs, \DESIGN{} improves performance by up to 2.5$\times$ (43\% geomean per app) over sequential execution and 2$\times$ (18\% geomean per app) over naively exploiting all parallelism, without requiring hardware changes.
\DESIGN{} also improves performance over hardware-partitioned GPUs~\cite{mig, mxgpu}, and \DESIGN{}'s benefits increase with reduced precision and as FLOPS scale, underscoring its importance given hardware scaling trends.

%% file: tables/conc_capability.tex
{\scriptsize
\begin{tabular}{cc|cc|}
\cline{3-4}
\multicolumn{2}{c|}{} &
  \multicolumn{2}{c|}{ \textbf{Operator Optimization Environment}} \\ \cline{3-4} 
\multicolumn{2}{c|}{\multirow{-2}{*}{}} &
  \multicolumn{1}{c|}{\textbf{Isolated}} &
  \textbf{Global} \\ \hline
\multicolumn{1}{|c|}{\cellcolor[HTML]{FFFFFF}} &
  \textbf{Static} &
  \multicolumn{1}{c|}{\begin{tabular}[c]{@{}c@{}}Current GPUs, \\ MIG/MxGPU~\cite{mig, mxgpu} \end{tabular}} &
  \begin{tabular}[c]{@{}c@{}} Rammer\cite{MaXie2020-rammer},\\ Elastic Kernels\cite{pai2013improving} \end{tabular} \\ \cline{2-4} 
\multicolumn{1}{|c|}{\multirow{-2}{*}{\textbf{\begin{tabular}[c]{@{}c@{}}Concurrency \\ Control Logic\end{tabular}}}} &
  \textbf{Dynamic} &
  \multicolumn{1}{c|}{ \begin{tabular}[c]{@{}c@{}} Queue/WF \\ schedulers \end{tabular}} &
  \textit{\textbf{\DESIGN{}}} \\ \hline
\end{tabular}%
}

%% file: 03_background_motivation.tex
\section{Background \& Motivation}
\label{sec:back}

\begin{figure}[tb!]
  \includegraphics[width=0.7\columnwidth, trim={3cm 11.25cm 13cm 6cm}]{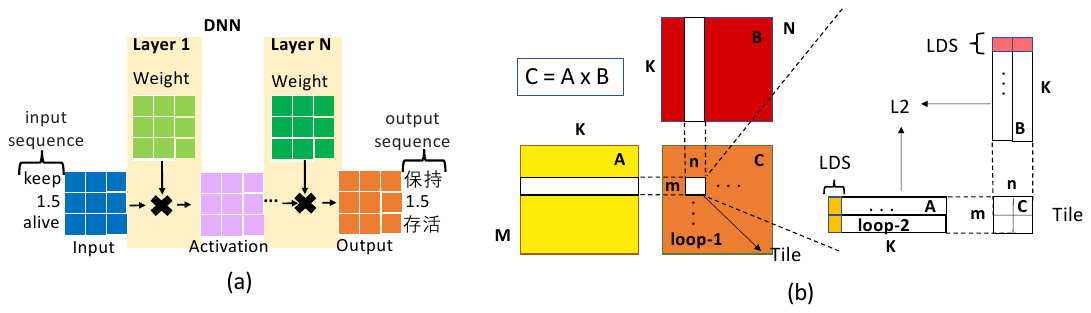}
  \centering
  \caption{(a) Toy DNN computation. (b) High-level GEMM implementation on a GPU.}
  \label{fig:sqnn_to_gemm}
\end{figure}

\subsection{GEMM: a prominent GPU operation}
\label{subsec:back-gemms}

\subsubsection{GEMM's dominance}
\label{subsubsec:back-gemms-time}

While GPUs run many different operations, they frequently execute highly parallel GEMM operations.
Furthermore, most of a DNN's execution manifests as GEMMs.
Figure~\ref{fig:sqnn_to_gemm}(a) shows a common DNN setup: DNNs have a series of layers, each of which executes as a GEMM between the input and the layer's weight matrix. DNNs also have non-GEMMs, including element-wise adds, and activations~\cite{ba2016layer} but they are often fused with preceding GEMMs via kernel fusion~\cite{ElHajjGomezLuna2016-klap, FousekFilipovivc2011-fuseGPUMap, SpringerWauligmann2017-fuseGPULang, WangLin2010-gpuKernelFusion} and tensor contractions~\cite{KimSukumaran2018-tensorContractGPU,KimSukumaran2019-tensorContractGPU, NelsonRivera2015-tensorContractGPU, ShiNiranjan2016-tensorContractGPU} to reduce memory traffic and kernel launch overheads.
Thus, GEMMs usually dominate DNN runtime~\cite{he2020newton, qin2020sigma, PatiAga2022-demystifying}.

\subsubsection{GEMM Operation}
\label{subsubsec:back-gemms-oper}

As shown in Figure~\ref{fig:sqnn_to_gemm}(b), a GEMM multiplies two input tensors $A$ and $B$ of size $MxK$ and $NxK$, respectively, to generate an output tensor $C$ of size $MxN$.
This involves $2*M*N*K$ floating point multiplies and adds.
The values of $M$, $N$ and $K$ are usually dictated by model hyperparameters such as layer width, batch-size, and/or input length (sequence length).
Additionally, the input tensors may be used transposed or non-transposed or both (e.g., transposed in forward propagation but non-transposed in back propagation).
We represent the transpose of $A$ and $B$ input tensors by $T1, T2$ (e.g., 1,0 implies only tensor $A$ is transposed).

\subsubsection{GEMM GPU Implementation}
\label{subsubsec:back-gemms-impl}

In GPU GEMM implementations $C$ is often blocked/tiled (\textit{Tile} in Figure~\ref{fig:sqnn_to_gemm}(b)) with each work group (WG) usually responsible for a single tile (loop 1).
Each thread in the WG multiplies and accumulates a row(s) with its respective column(s) within the innermost loop (loop 2).
These threads often leverage fast on-chip shared memory or LDS to store row/column data.
Several optimizations are usually applied, including executing a subset of WGs at a time (which impacts cache reuse), prefetching data from memory to the LDS, and coalescing.
Unlike other operations, applying these optimizations make GPU GEMM implementations quite complex, with hundreds of tunable features per size/transpose combination.
Thus, to improve performance, vendors rigorously tune implementations for GEMMs of different sizes, corresponding to different layer types or parameters~\cite{Tensile}.

\begin{figure}[tb]
\centering
\includegraphics[width=\columnwidth, trim={0cm 10cm 16cm 4cm}]{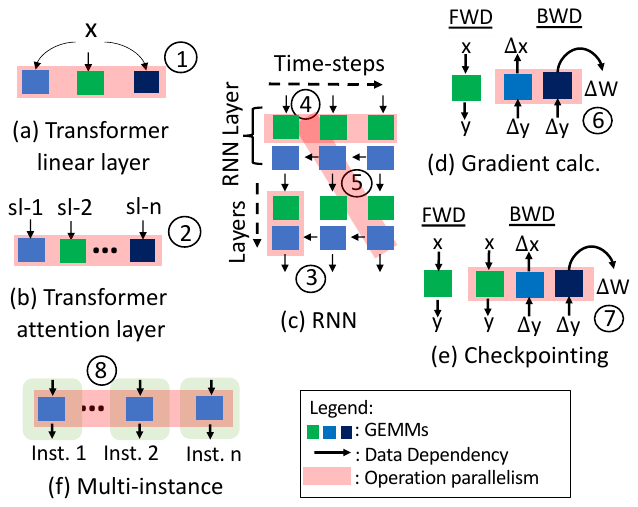}
  \vspace{-1ex}
  \caption{ML algorithms with independent operations.} 
  \label{fig:types_parallelism}
\end{figure}

\subsection{Important DNNs with GEMMs}
\label{subsec:bkg-sqnns}

Given their popularity~\cite{StateofAI22} and abundant parallelism, we focus on natural language processing (NLP)-based DNNs~\cite{DevlinChang18-bert,he2018streaming,VaswaniShazeer17-attention}, including \textbf{Recurrent Neural Networks (RNNs)} and \textbf{Transformers}.
However, \DESIGN{} also applies to other DNNs (Section~\ref{sec:disc}).
Table~\ref{tab:apps} lists the workloads we study.
RNNs process one token of the input at a time~\cite{ChoMerrienboer14-gru, HochreiterSchmidhuber1997-lstm, Rumelhart88RNN}.
The token processing manifest as one or more GEMM(s) and the sequential nature makes the input tensor to the GEMMs (in Figure~\ref{fig:sqnn_to_gemm}) small, with one of the dimensions equal to the input batch size.
Transformers use attention layers~\cite{BahdanauCho2015-nmt, VaswaniShazeer17-attention} to represent a token as the weighted sum of all other tokens in the input sequence.
Thus, they process all tokens of a sequence in parallel using an operation that manifests as a GEMM. However, each input in a batch must be processed independently as a separate GEMM.

\subsection{Scaling GPUs and Low Utilization GEMMs}
\label{subsec:back-util}

Both GPUs cores and their peak achievable FLOPS have scaled considerably. For example, between 2022 and 2023 FLOPS scaled by 4$\times$~\cite{mi210,mi300x}.
However, GPU utilization for applications like NLP-based DNNs often remains low.
GEMMs GPU utilization can be low when the input/output matrix sizes (Figure~\ref{fig:sqnn_to_gemm}(a)) are small.
This is common in DNNs (Section~\ref{subsec:bkg-sqnns}) due to their training/inference setup and/or algorithmic properties, including lower input batch sizes, short Transformer input sequences, and sequential RNN input token processing.
Reducing input batch sizes helps memory capacity requirements, improves convergence during training, and helps meet application deadlines during inference~\cite{jain2018dynamic}.
However, smaller batch sizes also limit matrix sizes, hurting utilization and throughput (e.g., only up to 23\% of TPU peak  throughput~\cite{JouppiYoung2017-tpu}).
Short Transformer input sequences (e.g., length 512 BERT attention GEMMs only achieve 25\% of peak throughput across vendors~\cite{IvanovDryden2021-transformersDataMove,PatiAga2022-demystifying}), and sequential RNN input token processing also limit matrix sizes (e.g., 2-30\% utilization~\cite{HolmesMawhirter2019-grnn,liu2019performance,ZhangRajbhandari2018-deepCPU,MaXie2020-rammer}).
Figure~\ref{fig:sqnn_to_gemm}(a)'s weight matrix can also be small: BERT GEMMs only achieve 40-50\% of peak FLOPs across GPU vendors~\cite{IvanovDryden2021-transformersDataMove, PatiAga2022-demystifying, ShoeybiPatwary2019-megatronlm, ZadehPoulos2019-dlTime}.
Larger models may slice matrices with tensor parallelism~\cite{megatron-github}, which reduces per-device memory capacity pressure but decreases their GEMM utilization.
In other work up to 90\% of ML-as-a-service (MLaaS) workloads also utilize GPUs poorly~\cite{WengXiao2022-mlaas, XiaoRen2020-antman, RecasensZhu2024-smallMLServing, ZhangElnikety2020-modelSwitching}.
Thus, ML workloads often do not utilize modern GPUs well and utilization trends will worsen with continued GPU FLOP scaling.

\subsection{Opportunities for GEMM Concurrency in DNNs}
\label{sec:char}

While individual DNN GEMMs have low GPU utilization, overall device utilization can be improved by concurrently executing multiple independent operations.
As shown in Figure~\ref{fig:types_parallelism}, DNNs have abundant opportunities to do so: they possess considerable operation parallelism from their model architecture.
These include independent query/key/value generation in the linear layers, and independent (batched) attention computations for unique sequence length (SL) inputs in Transformers (\circled{1} and \circled{2} in Figure~\ref{fig:types_parallelism}), respectively. Note, the latter is required to avoid padding of sequences to the maximum length and avoid extraneous computations~\cite{mlperf-training-nvidia-v20}.
Similarly, independent input processing in the time dimension and hidden state processing across layers in RNNs introduce operation parallelism (\circled{3}, \circled{4}, \circled{5}).
Training algorithms also have additional parallelism opportunities that apply to all DNNs (e.g., CNNs, Recommendation). These include independent weight and input gradient calculations during back-propagation (\circled{6}) and activation recomputing due to checkpointing (\circled{7}).
Finally, while not applicable during training (due to large memory capacity requirements), multiple DNN inference instances (\circled{8}) are deployed on the same GPU in production environments which provides additional concurrency opportunities~\cite{ChoiKim2021-lazyBatching, ChoiRhu2020-prema, GaoYu2018-batchmaker, JouppiYoon2021-tpuv4, KeGupta2020-recNMP, multi-instance-2, multi-instance-3, multi-instance-1, YehSinclair2021-lax}.

\begin{figure}[tb!]
  \centering
  \includegraphics[scale=0.46, trim={0cm 17cm 0cm 0cm}]{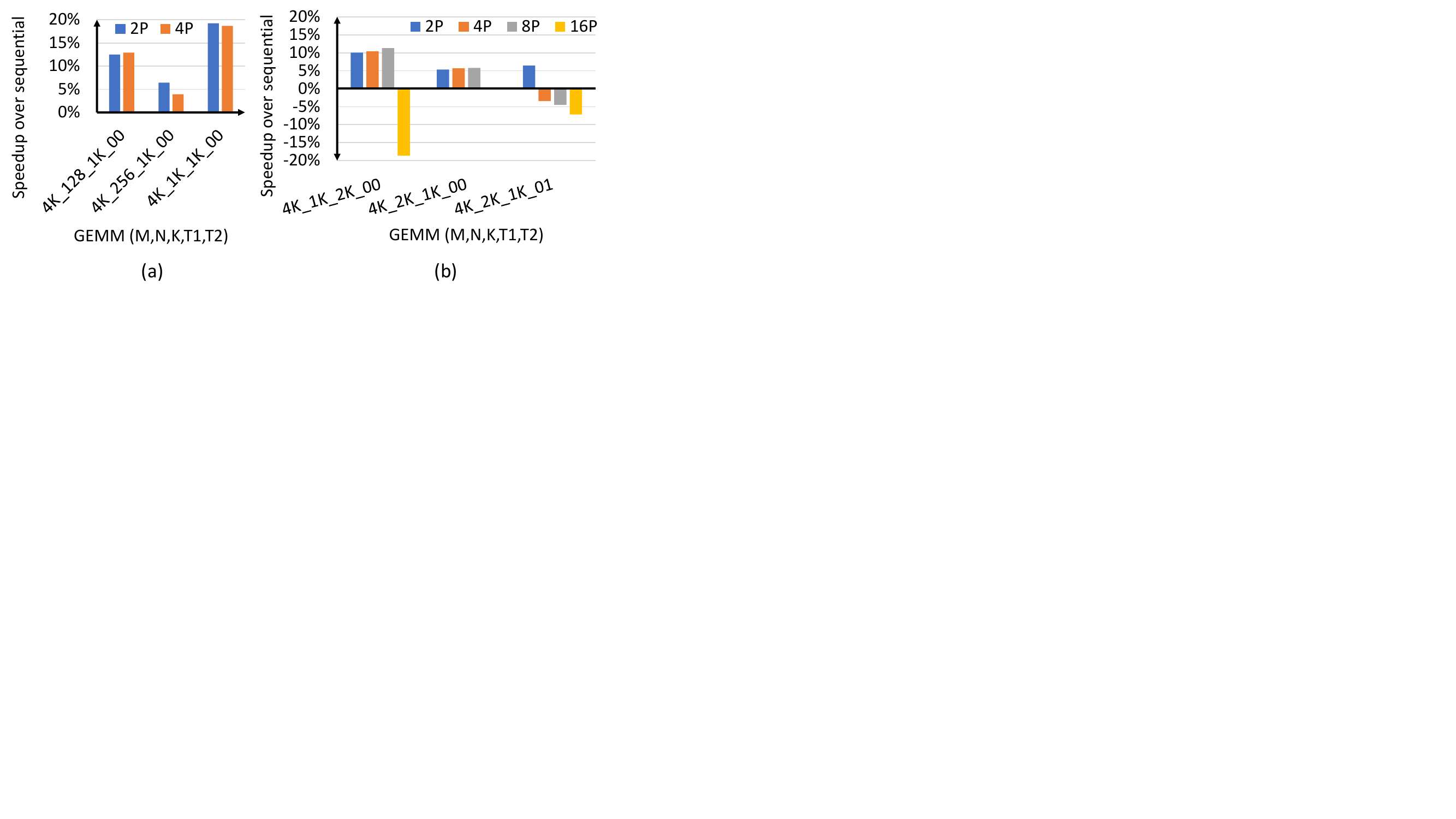}
  \caption{(a) GEMM sizes with fewer FLOPs benefit less from concurrency (b) GEMM sizes with the same FLOPs can have different concurrency behavior. X-axis represent GEMMs as M_N_K_T1_T2 and GEMM FLOPs are calculated as 2*M*N*K.}
  \label{fig:motiv}
\end{figure}

\subsection{Sub-optimal GEMM Concurrency in GPUs}
\label{sec:example_chal}

While there are abundant opportunities to concurrently execute low utilization GEMMs, naively executing them concurrently often provides small performance improvements on GPUs.
Figure~\ref{fig:motiv} illustrates this with a few examples.
First, Figure~\ref{fig:motiv}(a) shows the speedups when concurrently executing two and four independent GEMMs (IG=2, 4) over sequentially executing them.
Figure~\ref{fig:motiv}(a) also evaluates this for multiple GEMM sizes, with the size of GEMMs (particularly the $N$ dimension) increasing from left to right.
While the largest GEMMs achieve $\approx19$\% speedup over their sequential execution, the smaller ones (with fewer FLOPs) achieve much smaller speedups. Thus, counter-intuitively, GEMMs with smaller compute requirements benefit less from concurrent execution.

Figure~\ref{fig:motiv}(b) studies GEMMs with the same FLOPs but different input tensor shapes (the first two) or transposes (the last two), for IG=2,4,8,16.
The first two cases speedups' over sequential execution are similar or slightly increase as concurrency degree increases from 2 to 8 IGs. 
However, for 16 IGs performance degrades for \textit{4k\_1k\_2k\_00}.
For \textit{4k\_2k\_1k\_01}, which has a transposed input, B, tensor, performance degrades for all IGs beyond two.
Thus GEMMs, even those with similar compute requirements, can have very different concurrency behavior and do not always benefit from concurrency.
In Section~\ref{sec:challs} we further study and identify challenges in current GPUs that result in these behaviors.

%% file: 04_challenges.tex
\section{Challenges with GEMM Concurrency}
\label{sec:challs}

Next we examine how Section~\ref{sec:example_chal}'s examples reinforce Table~\ref{tab:capability_overview}'s two key challenges with leveraging GPU concurrency.

\begin{figure}[tb!]
  \includegraphics[scale=0.7]{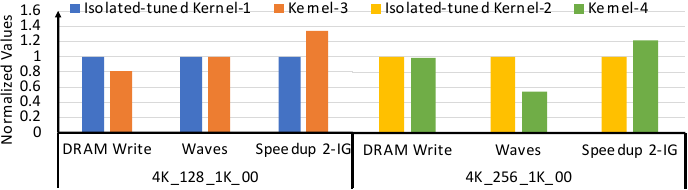}
  \centering
  \caption{GEMM behavior with different kernel implementations. Kernels-1 and -2 are the GEMMs' isolated tuned kernels; Kernels-3 and -4 are alternate implementations with smaller memory traffic and fewer WG waves, respectively.}
  \label{fig:tuning_motiv}
\end{figure}

\subsection{Isolation-tuned kernel implementations}
\label{subsec:challs-lib}

Figure~\ref{fig:motiv} showed that the GEMM with the most FLOPs benefited more from concurrency.
Besides size, these GEMMs have different kernel implementations (e.g., the largest GEMM has the largest tile size, among other differences).
A GEMM's kernel implementation involves tens of features that are tuned to improve its \textit{isolated} GPU execution (Section~\ref{subsubsec:back-gemms-impl}).
As a GEMM's hardware requirements differ based on its input (size, shape, transpose), they also prefer unique kernel features for maximum performance: the 410 GEMM sizes we study (Section~\ref{sec:meth}) chose 291 unique kernel implementations.

Kernel implementations also have a significant impact on concurrent performance: a larger tile size reduces the number of WGs a GEMM executes but increases the extent of LDS data reuse.
Features such as coalescing limit global memory traffic while also increasing register/LDS requirements and decreasing per compute unit (CU) occupancy. 
The WG count and occupancy impact how concurrent GEMMs share CU resources, while data reuse and total memory traffic impact how they share the cache/memory bandwidth.
Similarly, every other feature has a unique trade-off.

Figure~\ref{fig:tuning_motiv} evaluates the two smaller FLOPs GEMMs from Figure~\ref{fig:motiv} using alternate, more concurrency-amenable kernels.
\textit{Isolation-tuned} Kernel-1 and Kernel-2 are tuned for the GEMM's performance in isolation, as in BLAS libraries.
Compared to Kernel-1, Kernel-3 improves both LDS reuse (via larger tile size) and the kernels' accesses to the LDS (via padding and prefetching).
Consequently, this reduces the \textit{4k\_128\_1K\_00} GEMM's global memory accesses and improves its two concurrent independent GEMMs performance by 1.34$\times$.
Conversely, Kernel-4 slightly increases the number of WGs (smaller tile size) and reduces LDS requirements (via less coalescing) compared to Kernel-2, which improves the GEMM's CU occupancy by 2$\times$ for \textit{4k\_256\_1K\_00} and reduces the number of waves (set of WGs a kernel simultaneously executes on a GPU).
This improves the GEMMs' overlap and increases two concurrent GEMMs speedup by 1.22$\times$.

Overall, these exemplar results show that considering the \textit{global} resource environments for kernel implementations, based on the operations executing concurrently,
can improve performance.
However, there are two challenges in realizing them: (a) as shown in Figure~\ref{fig:tuning_motiv} GEMMs have different (e.g., memory, compute) bottlenecks depending on the input properties and must optimize for different resources and (b) there are several kernel features, each with a unique trade-off, that can be tweaked to optimize for the bottlenecked resource -- and similar to the baseline BLAS libraries, these will differ for each GEMM input.
Thus, manually identifying such alternative implementations is challenging.
Therefore, we need a method to systematically identify \textit{globally} optimized kernels for many different GEMMs.

\begin{figure}[tb!]
  \centering
  \includegraphics[scale=0.5]{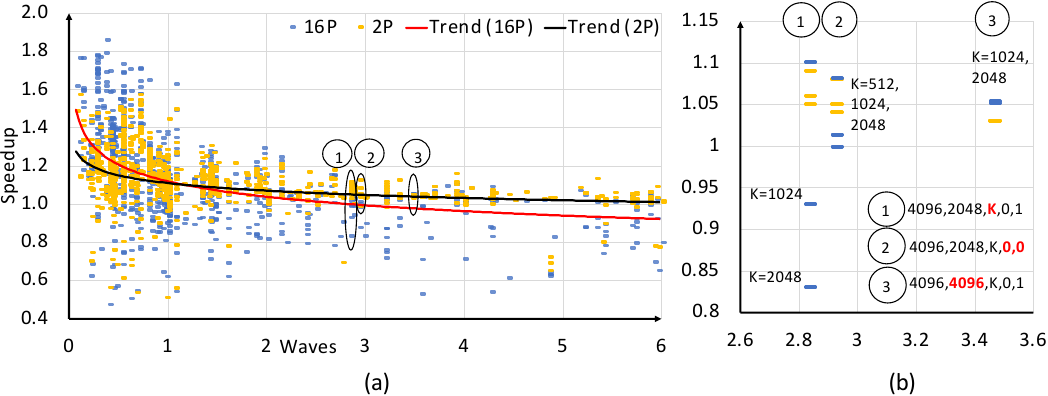}
  \vspace{-3ex}
  \caption{(a) Speedups over sequential execution for 2 \& 16 concurrent GEMMs (2P \& 16P)  versus the \#waves in their isolated execution. (b) Speedups of GEMMs with fixed \#waves but with varying K, input shape, or transpose.}
  \label{fig:conc_factors}
\end{figure}

\subsection{Static concurrency control}
\label{subsec:challs-amount}

Figure~\ref{fig:conc_factors}(a) examines how the 410 studied GEMMs (Section~\ref{sec:meth}) perform when running two and 16 concurrent, independent GEMMs.
The x-axis shows the number of waves used by the GEMM kernels.
In general, fewer wave GEMMs have better concurrency behavior: higher 2-IG speedups and benefits with higher concurrency degrees (e.g., for 16-IG).
This matches our earlier observation (Section~\ref{subsec:challs-lib}) that smaller/fewer waves enable better overlap/sharing of CUs.
However, the behavior varies significantly for GEMMs with similar waves.
We highlight this using examples \circled{1}, \circled{2}, and \circled{3}, zoomed in on Figure~\ref{fig:conc_factors}(b).
\circled{1} compares concurrently executing GEMMs with the same $M, N, T1$, $T2$, and number of waves, but different $K$ dimensions.
Their performance varies considerably; for example, performance degrades at $K$ of 1024 and 2048.
The summation dimension ($K$) determines the amount of work performed and data read per thread and per WG.
Our profiling of isolated GEMM execution\footnote{AMD and NVIDIA GPUs currently do not support performance counter monitoring with concurrent kernels.} shows that increasing $K$ also increases the memory reads-to-input matrix size ratio, implying larger $K$ GEMMs are more prone to Last-Level Cache (LLC) and memory bandwidth contention.

Similarly the transpose combination ($T1$,$T2$) determines the GEMM input tensors' memory layout and thus its memory access pattern.
Certain transpose combinations have better data locality and improve cache/bandwidth sharing during concurrency.
\circled{2} in Figure~\ref{fig:conc_factors}(b) compares concurrently executing GEMMS with the same GEMM dimensions and similar waves as \circled{1}, but a different ($0$,$0$) transpose.
Unlike \circled{1}, these GEMMs do not see performance degradation. 
Finally, the shape of tensors also dictates behavior.
Generally, similar-sized inputs ($M \approx N$) indicate that input rows and columns have similar cache reuse.
Therefore, \circled{3}, which has similarly-sized inputs but larger GEMMs with more waves, also does not see \circled{1}'s performance degradation.

Across the 410 GEMMs in Figure~\ref{fig:conc_factors}(a) there are many such varied behaviors.
Whether GEMM concurrency is beneficial is dictated by a combination of input sizes, tensor shapes, layout, and kernel implementations -- not all of which are known statically.
\textbf{Furthermore, these concurrency benefits cannot be determined via simple heuristics and require profiling}.
Offline profiling could potentially identify the right amount of concurrency to exploit in every intra-application case. However, profiling at runtime to account for dynamic inputs and concurrent applications can add significant overheads and diminish concurrency benefits.
Thus, GPUs need lightweight, \textit{dynamic} logic to manage concurrency.

%% file: 05_proposal.tex
\section{\DESIGN{} Design}
\label{sec:prop}

\begin{figure}[tb!]
  \centering
  \vspace{-3ex}
  \includegraphics[scale=1.1]{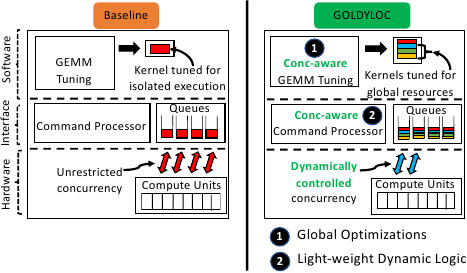}
  \caption{\DESIGN{} overview and baseline comparison.}
  \label{fig:design-overview}
  \vspace{-3ex}
\end{figure}

\subsection{Overview}
\label{sec:prop-overview}

Figure~\ref{fig:design-overview} depicts the baseline system (left) and \DESIGN{} (right).
We only show system components that \DESIGN{} affects.
In the baseline, there is a one-time GEMM library tuning for a given GPU such that, for a given GEMM size, at runtime the library returns a kernel optimized for its \textit{isolated} execution (Section~\ref{subsec:challs-lib}).
At runtime the command processor (CP), an embedded programmable microprocessor within the GPU which acts as the interface between the software and hardware~\cite{LeBeaneHamidouche2018-cpNet, LeBeanePotter2016-taskQueue}, schedules as many independent GPU kernels as possible given available resources~\cite{OtternessAnderson2021-amdSched, puthoor2018oversubscribed}.
This parallelism is either exposed by programmers via streams/queues \textit{statically}~\cite{nvidia-stream, nvidia-stream2} and/or from multiple processes.
In Figure~\ref{fig:design-overview}, the CP may schedule all four available GEMMs concurrently, each using an isolation-tuned kernel (four red arrows).

\textbf{\DESIGN{}} (Figure~\ref{fig:design-overview}, right) redesigns GPU libraries and runtime to add \textit{concurrency awareness} to the system.
Similar to the baseline, \DESIGN{} requires a one-time tuning of the GEMM library for a given GPU.
However, \DESIGN{} enhances the tuning methodology such that for a given GEMM size, at runtime, the library returns a kernel optimized for isolated execution and also kernels which are \textbf{\textit{globally optimized}} (GO-Kernels) for multiple concurrency degrees (CDs, i.e., number of concurrent GEMMs, Section~\ref{sec:prop-tuning}).
\DESIGN{} further programs the CP with a \textbf{\textit{lightweight dynamic logic}} to control the amount of concurrency on the GPU (Section~\ref{sec:prop-strategy-predictor}).
At runtime, given a set of independent GEMMs and their globally optimized kernels, the CP predicts a performant CD and schedules those many GEMMs with appropriate kernels.
For example, in Figure~\ref{fig:design-overview}, the CP dynamically predicts and schedules two of the four available GEMMs with kernels globally optimized for a CD of two (two blue arrows).
Thus, \DESIGN{} \textit{dynamically} selects and executes concurrent GEMMs which can improve overall performance, with kernels optimized for a \textit{global} shared resource environment.

\begin{figure*}[tb!]
  \begin{minipage}[b]{\linewidth}
  \centering
  \subfloat[][]{
     	\includegraphics[scale=0.58]{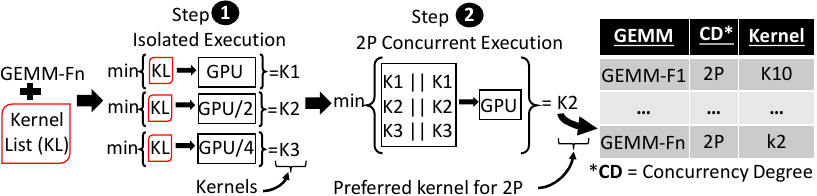}
    	\label{fig:design-tuning-1}
     }
     \vspace{-0.01cm}
     \subfloat[][]{
		 \includegraphics[scale=0.46]{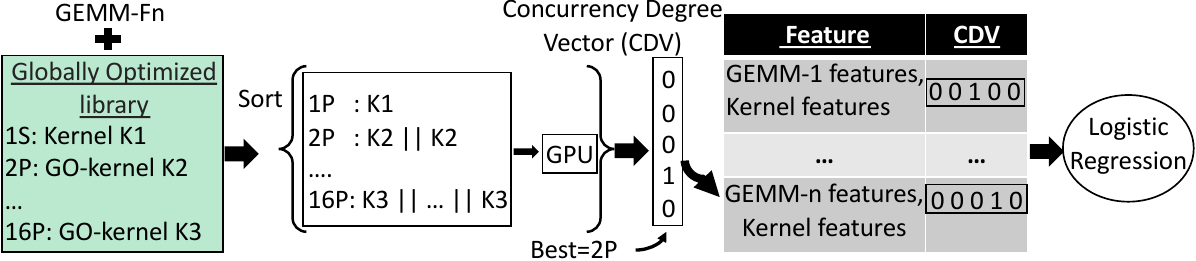}
    	\label{fig:design-predictor-process}
     }
  \end{minipage}
  \vspace{-2ex}
  \caption{(a) \DESIGN{}'s tuning methodology for a single GEMM for concurrency degree = 2P. (b) Identifying optimal concurrency degree for a single GEMM feature, and taming its overhead using a logistic regression-based model.}
\end{figure*}

\subsection{Globally optimized (GO) GEMM kernels}
\label{sec:prop-tuning}
Concurrently executing GEMMs with kernels tuned for isolated execution, as in the baseline, is suboptimal (Section~\ref{subsec:challs-lib}) and may hurt performance (Figure~\ref{fig:conc_factors}).
The baseline's rigorous benchmarking minimizes a kernel's latency assuming all GPU resources are available for a single GEMM.
This leads to kernels that may end up hoarding resources that must be shared during concurrent executions (e.g., the isolation tuned Kernel-1 is cache/memory bandwidth-heavy, while Kernel-2 is CU-heavy).
Therefore, GPUs must use kernels that are globally optimized (GO) for the available (shared) resources (e.g., Kernel-3 and Kernel-4 which limit the respective GEMMs' bandwidth and CU usage, respectively).
This requires identifying, for each given GEMM, which resources must be optimized for, and which kernel feature(s) to focus on to achieve that.
\DESIGN{} identifies such kernel implementations by augmenting the tuning process to include \textbf{resource constraints} (RCs).
Executing GEMMs with RCs emulates a concurrent environment where resources are shared, and thus limited.
Thus, tuning the kernel for each GEMM in such RC environments (Section~\ref{sec:prop-resource-constrained-tuning}) can help automatically identify the features optimized for the bottlenecked resource.

\subsubsection{Resource-constrained (RC) tuning}
\label{sec:prop-resource-constrained-tuning}

When incorporating RCs into tuning GPU kernel implementations, we must consider: which resources to focus on and how to augment tuning?

The most pertinent GPU resources are: CUs, cache, registers, LDS, and memory bandwidth. 
Although GPU configurations can be modified
to limit a kernel's on-chip resources (e.g., CUs, cache, LDS)~\cite{mig, OtternessAnderson2021-amdSched}, limiting memory bandwidth is more difficult.
Sophisticated data placement (e.g., over a subset of memory channels) adds significant software complexity.
Moreover, while tweaking memory frequency is possible, it may lead to lower access latency that may not be representative of access latency during concurrent execution.
Thus, we focus on constraining CU count and LLC size.
We create two 
RC configurations in addition to baseline GPU configuration (GPU): GPU/2 (halves \#CUs and LLC size) and GPU/4 (quarters \#CUs and LLC size).
We selected these based on available parallelism (or concurrency degree, CD) and empirical results which show little benefit from stricter RCs (Section~\ref{subsec:disc-rcs}). 

Figure~\ref{fig:design-tuning-1} shows how \DESIGN{} tunes for a given GEMM (GEMM-Fn).
The baseline tuning process rigorously benchmarks the available GPU kernels (Kernel List (KL)) on a resource-unconstrained GPU configuration.
Our tuning process also examines GPU/2 and GPU/4 (Step \circled{1}).
Next, using the set of most efficient kernels from Step \circled{1}, we benchmark concurrent execution for each CD of interest (e.g., 2P, 4P) (Step \circled{2}).
We benchmark kernels from all three RC configurations for all CDs.
For example, for CD=2P we benchmark kernels most efficient for GPU, GPU/2, and GPU/4. 
The kernel with the smallest runtime is preferred for the given GEMM and CD (K2 in Figure~\ref{fig:design-tuning-1}).
It is possible that a kernel tuned for isolated execution (RC=GPU) is also preferred for concurrency.
This happens if the GEMM is bound by a resource during its isolated execution and already selects the appropriate kernel to use that resource, requiring no further RC-tuning.
For example, very large compute-bound GEMMs often use kernels that limit the total WG and wave count.
This is also possible for small GEMMs at low CD which already have sufficient overlap and few waves (e.g., GEMMs with 0.5 waves will not benefit further from 0.25 waves). 
To reduce the benchmarking cost in Step \circled{2}, we also propose using similarity analysis to determine the RC configs preferred by GEMMs using exhaustive profiling of a subset of GEMMs (discussed in Section~\ref{sec:prop-taming-tuning-overhead}).

\begin{table}[tb!]
  \centering
  \resizebox{\columnwidth}{!}{%
    \input{tables/acronyms}
  }
  \caption{\DESIGN{} Acronyms}
  \label{tab:acronym}
\end{table}

\subsubsection{Globally optimized GEMM library}
\label{sec:prop-concur-aware-GEMM-library}

The baseline \textit{GEMM library} has GEMM inputs and associated GPU kernels optimized for isolated execution.
\DESIGN{} augments this library: during runtime each GEMM also returns pointers to globally optimized (GO) kernels efficient for the global resource environment per CD (\circled{1}). We discuss this further in Section~\ref{sec:prop-design-integration}.

\subsection{Dynamic logic for concurrency control}
\label{sec:prop-strategy-predictor}

Baseline GPUs statically control concurrency within applications, without  knowledge about dynamic input sizes or number of processes.
As observed in Section~\ref{subsec:challs-amount}, this can degrade performance since not all concurrency is beneficial, even when using GO kernels.
Moreover, while dynamic control is important, determining the appropriate amount of concurrency at runtime is challenging.
It depends on a combination of factors (GEMMs' tensor size, shape, and layout as well as kernel implementation (Figure~\ref{fig:conc_factors}(b)) and requires profiling which can add significant overheads at runtime.
To overcome this, \DESIGN{} uses one-time offline profiling of a subset of GEMMs and trains a lightweight predictor to determine the appropriate CD to execute at runtime.

\noindent
\textbf{Offline profiling \& predictor dataset}: Figure~\ref{fig:design-predictor-process} depicts \DESIGN{}'s offline profiling, which identifies the appropriate CD for a GEMM and creates the dataset used to train the predictor.
For a given GEMM \DESIGN{} benchmarks the kernels identified by the GO GEMM library with their associated CD (e.g., 2P uses GO K2).
Amongst all possible CDs, it associates this GEMM with the CD that delivers the most speedup over its corresponding serial execution.
Increasing the number of concurrent GEMMs up to this CD often improves performance but further increases either provide no further improvement or degrade performance.
Thus, the final executed CD should be the minimum of this preferred CD and the available GEMMs.

Based on our observations in Section~\ref{subsec:challs-amount} 
\DESIGN{} uses GEMM dimensions and its per-CD kernels' (\#WGs, occupancy, and \#waves) as the predictor's input features as they capture all input, implementation, and underlying GPU's hardware properties.
\#WGs is a function of output size ($M$$\times$$N$) and determines total parallelism within the GEMM.
Occupancy accounts for each WG's resource requirements, hardware resources per CU, and potential L1 cache contention.
Wave count incorporates total CU count in hardware, kernel tile size, and potential for overlap.
Finally, size (specifically, $K$) and shape ($M$, $N$) provide information on memory contention.
We also considered other kernel features (e.g., grid size, LDS/register size) and performance data, but they provided minimal accuracy improvements. 

\noindent
\textbf{Logistic regression model details}: 
To compare different CD's relative benefits \DESIGN{} trains a multi-class (one class per CD) logistic regression model~\cite{carroll1993robustness, hosmer1997comparison, sklearn-multiclass-log-reg}.
Logistic regression is a good choice as GEMMs have multiple input features with near-linear relationships with concurrency benefits (e.g., speedup drops with increasing $K$) and because it generates a multi-class output (either no concurrency or CD of 2, 4, 8, 16).
The predictor calculates the probability of preferring one CD over the rest (one-vs-rest, OvR) and predicts the appropriate CD, including no concurrency. Training it 
fits (learns the weights of) Equation~\ref{eq:logRegr}:

\begin{equation}
  P = \frac{e^{X \times W}}{ \sum_i^C e^{X \times W_i}}
  \label{eq:logRegr}
\end{equation}

\noindent
where $P$ is the probability vector to select one CD over the rest,
$X$ ($x_1, x_2,.., x_n$) are input features, $W$ is the weight matrix and $C$ is the possible CD count. 

\begin{figure}[tb!]
    \centering
    \includegraphics[width=0.8\columnwidth, trim={1cm 11cm 6cm 3.1cm}]{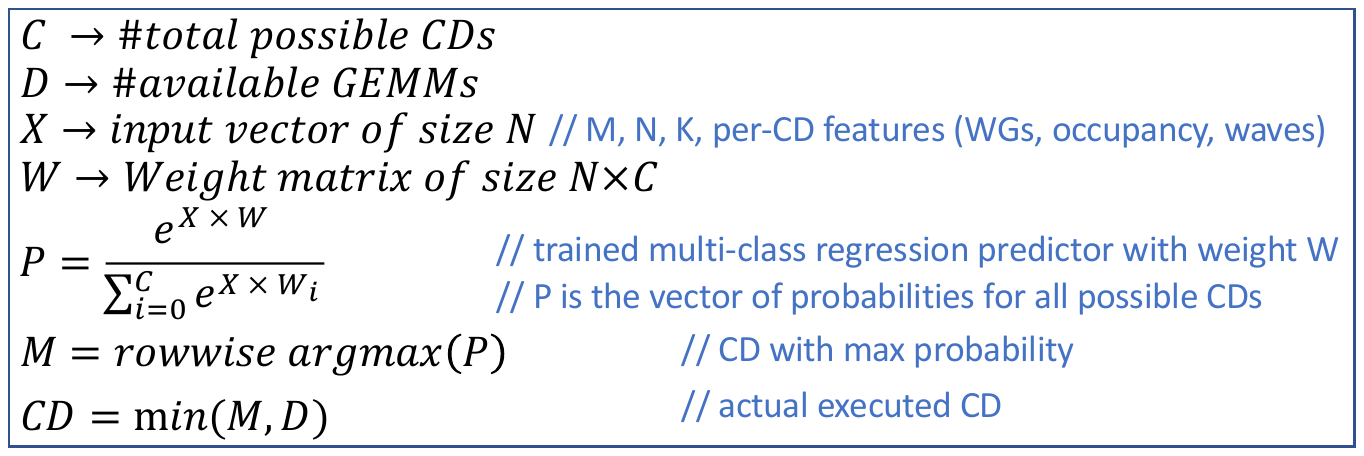}
    \vspace{-1ex}
    \caption{\DESIGN{}'s dynamic logic.}
    \vspace{-3ex}
    \label{fig:sp_alg}
\end{figure}

We train the predictor on a dataset created from offline profiling.
In the training dataset all GEMMs' features are mapped to their preferred CD (Figure~\ref{fig:design-predictor-process}'s table).
To create a more exhaustive dataset we include additional GEMMs beyond the evaluated workloads, for a total of 1072 GEMMs.
We apply min-max normalization to normalize the dataset feature values.
\DESIGN{} trains the model offline once per GPU (accuracy discussed in Section~\ref{subsec:eval-accuracy}) using 90\% and 10\% samples for training and testing, respectively.
After training, it predicts the appropriate CD (1S, 2P, 4P, 8P, or 16P, Figure~\ref{fig:sp_alg}).
Given the queued GEMMs' feature vector, $X$, and learned weights, $W$, it calculates the probability to choose each possible CD (total $C$) with Equation~\ref{eq:logRegr} and selects the one with maximum probability. 
The final chosen CD is the minimum of the predicted CD and available GEMMs. 
Figure~\ref{fig:design-tuning-predictor} shows how \DESIGN{} incorporates this predictor into the GPU CP (discussed further in Section~\ref{sec:prop-design-integration}).

\subsection{Integrating \DESIGN{} into GPU's CP}
\label{sec:prop-design-integration}
\noindent
\textbf{Kernel-packet Extensions}: To schedule a GEMM on a GPU, CPUs enqueue a \textit{kernel packet}~\cite{amd-kernel-object} in the CP's queues on that GPU.
This packet
includes a pointer to the kernel object (KO) that is invoked to execute the GEMM, along with its associated metadata such as the kernel's input arguments and features (e.g., WG size).
The packet also includes  
additional header, setup, and reserved bytes.
Since identifying the appropriate GO kernel, and thus the appropriate KO, for a given GEMM requires dynamic information about available parallelism and input sizes, a kernel packet cannot be pre-mapped to a single KO.
Instead, \DESIGN{} extends kernel packets 
to include a map of KO pointers and metadata for each GO kernel (max three per GEMM from the three RC configurations) from the GO library (Section~\ref{sec:prop-concur-aware-GEMM-library}).
These extensions add a little overhead, but since KOs are relatively small and only in CP memory until dispatch completes, the packets still fit in the CP's memory. 

\begin{figure}[tb!]
    \centering
    \includegraphics[scale=0.54]{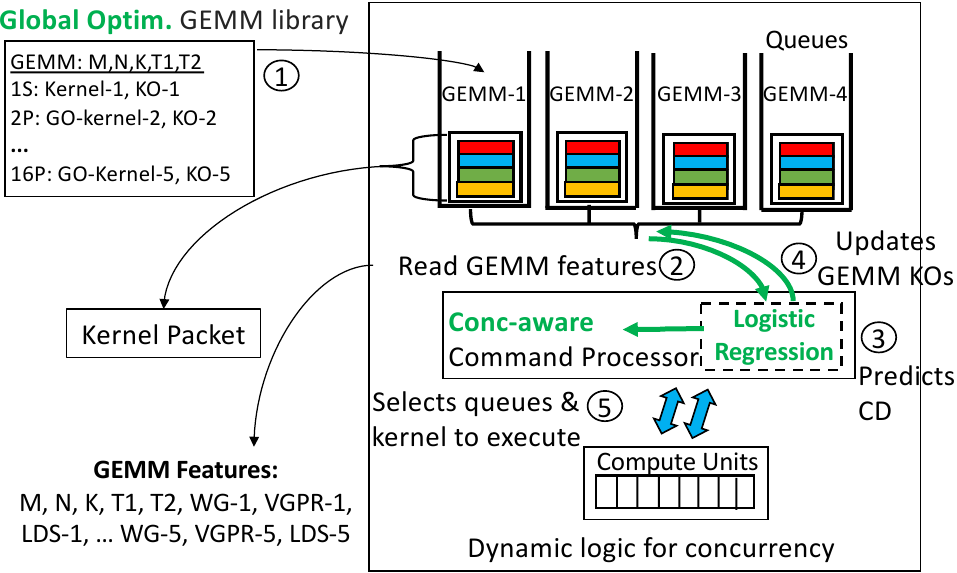}
    \caption{\DESIGN{} GEMM library and dynamic logic.  WG-$i$, LDS-$i$, and VGPR-$i$ represent the WG, LDS, and vector register usage for different concurrency-tuned kernel implementations (CDs) for a given GEMM, respectively}.
    \label{fig:design-tuning-predictor}
    \vspace{-2ex}
\end{figure}

\noindent
\textbf{Command Processor Extensions}:
At runtime, current GPU CPs inspect all available software queues (streams) and their kernels to schedule as many independent kernels from separate queues as resources permit~\cite{OtternessAnderson2021-amdSched, puthoor2018oversubscribed}.
Thus, the CP is well suited to dynamically control the amount of concurrency.
\DESIGN{} programs the CP to inspect the kernel packets at the head of all active queues (\circled{2} in Figure~\ref{fig:design-tuning-predictor}) for available independent GEMMs that could execute concurrently.
This includes (a) checking if the kernels are GEMMs or non-GEMMs, (b) if there are multiple GEMMs, reading the necessary features from queued packets,  
and (c) calculating the remaining features (occupancy and waves) needed for prediction.
The CP performs these operations each time a queue's head changes -- when a kernel finishes dispatching its WGs or when new work is enqueued.
CP functionality is unchanged if it detects a single GEMM and/or non-GEMMs.
For multiple GEMMs, given the number and features of the GEMMs, the CP predicts the appropriate CD (\circled{3} in Figure~\ref{fig:design-tuning-predictor}); 
both the right (set of) GEMM(s) and how many GEMMs to execute concurrently.
Finally, the CP updates the packet contents of these GEMMs, located at the queue heads, to point to the KO corresponding to the GO kernel for the predicted CD (\circled{4} in Figure~\ref{fig:design-tuning-predictor}) which are then executed on the GPU (\circled{5} in Figure~\ref{fig:design-tuning-predictor}).

%% file: tables/acronyms.tex
{\scriptsize
\resizebox{\columnwidth}{!}{
\begin{tabular}{|c|c|c|c|}
\hline
{\color[HTML]{333333} \textbf{Acronym}} & {\color[HTML]{333333} \textbf{Definition}}           & {\color[HTML]{333333} \textbf{Acronym}} & {\color[HTML]{333333} \textbf{Definition}} \\ \hline
{\color[HTML]{333333} CD}               & {\color[HTML]{333333} Concurrency Degree}            & {\color[HTML]{333333} CP}               & {\color[HTML]{333333} Command Processor}   \\ \hline
{\color[HTML]{333333} RC}               & {\color[HTML]{333333} Resource Constraint}           & {\color[HTML]{333333} CU}               & {\color[HTML]{333333} Compute Unit}        \\ \hline
{\color[HTML]{333333} nP}               & {\color[HTML]{333333} $n$ Parallel GEMMs}      & {\color[HTML]{333333} KO}               & {\color[HTML]{333333} Kernel Object}       \\ \hline
{\color[HTML]{333333} GO}              & {\color[HTML]{333333} Globally Optimized} & {\color[HTML]{333333} LLC}              & {\color[HTML]{333333} Last Level Cache}    \\ \hline
\end{tabular}%
}
}

%% file: 06_meth.tex
\section{Methodology}
\label{sec:meth}

\subsection{System Setup}
\label{subsec:meth-setup}

We evaluate \DESIGN{} with AMD
ROCm\texttrademark\ platform because it has a high performance, open-source BLAS tuning framework. 
Specifically, we extend AMD's ROCm 4.1~\cite{rocm} libraries by using Tensile~\cite{Tensile} for tuning and rocBLAS~\cite{rocblas} to build the custom BLAS libraries.
Both the tuner and the library utilize Matrix Core Engines~\cite{amd-cdna-arch}.
Moreover, we use an AMD Ryzen\texttrademark\ Threadripper\texttrademark\ CPU~\cite{threadripper} and an AMD Instinct\texttrademark\ MI100 GPU~\cite{mi100} with 32GB of HBM2~\cite{hbm2}.
We calibrated this system's baseline performance and found it was similar to other commercial systems and prior work~\cite{IvanovDryden2021-transformersDataMove}: all had similar FLOPS relative to the peaks, and 90\% of all studied GEMMs had differences within -12\% to +10\%.

\subsection{Applications and GEMMs Studied}
\label{subsec:meth-apps}

To evaluate \DESIGN{} we use 410 GEMMs (Table~\ref{tab:apps}) from forward and backward passes of state-of-the-art RNNs and Transformers while varying their batch and token sizes ("Input Params" in Table~\ref{tab:apps}).
Similar to modern datacenters deployments~\cite{KatoLakshmanan2011-timeGraph} and recent work~\cite{ChoiKim2021-lazyBatching,ChoiRhu2020-prema,GaoYu2018-batchmaker,pats}, we evaluate independent GEMMs both within and across networks for multi-instance inference deployments (Section~\ref{sec:char}): 2, 4, 8, and 16 instances (there were diminishing returns beyond 16).
To create a more representative dataset we include additional GEMMs (1072 total). 
The GEMM's ranges are: 32K-168M for output size (M*N, dictates parallelism), and 64-20K for K dimension (dictates data per thread/WG). They represent a wide variety of memory and compute-bound behavior; ops/byte (dictates memory-boundedness) ranges from 28-1400.
We examine both full and half precision GEMMs.
Finally, we also study concurrent strided batched-GEMM (B-GEMMs) from Transformer Attention layers (with over 40 combination of different SLs).

\subsection{Measurement}
\label{subsec:meth-measure}
For GO kernel tuning and profiling for dynamic predictor datasets
(Section~\ref{sec:prop}), we execute GEMMs with different RCs, CDs (via GPU streams), and kernels.
To average out queuing delays in concurrent setups we execute GEMMs back-to-back on the same stream multiple times.
We measure runtimes using rocProf~\cite{rocprof-profiler}.

\begin{table}[tb!]
  \centering
    \resizebox{\columnwidth}{!}{%
  \input{tables/bmks-new}
  }
  \caption{Benchmarks with hyperparameters and inputs.}
  \label{tab:apps}
\end{table}

\subsection{\DESIGN{} Performance Measurement}
\label{subsec:meth-perf}

\subsubsection{Globally Optimized (GO)-Kernels}
\label{subsubsec:meth-perf-go}
We modify the Tensile~\cite{Tensile} tuning infrastructure to create a custom globally optimized library (Section~\ref{sec:prop-tuning}).
Sequential GEMM applications use the baseline library.
We create two binaries of the concurrent GEMM application, each linked to the baseline or GO library. To evaluate GO-Kernels, for each GEMM size, we find the speedup of 
the concurrent binaries (with different CDs) over the sequential run of the GEMM.

\subsubsection{\DESIGN{}}
\label{subsubsec:meth-perf-goldyloc}
Although the dynamic control logic can be implemented in existing GPUs by reprogramming the CP, GPU vendors have not disclosed an API~\cite{LeBeaneHamidouche2018-cpNet, LeBeanePotter2016-taskQueue, YehSinclair2021-lax}.
We also implemented our changes in gem5's CP~\cite{BruceAkram2021-gem5art, GutierrezBeckmann2018-gem5GPU, LowePowerAhmad2020-gem520, RoartySinclair2020-gem5GPU} but like prior work found its performance trends did not match real hardware~\cite{JamiesonChandrashekar2022-gap, RamadasKouchekinia2023-gap}.
Thus, we evaluate \DESIGN by measuring the runtime of concurrent GEMMs with CD predicted by the dynamic logic (using the custom GO library) on real hardware and add our CP modification overheads.

We model the CP's dynamic detection, prediction, and selection (Section~\ref{sec:prop-design-integration}) latency.
This includes the CP's kernel packet reads and writes from queues and logistic regression model execution. 
We model the CP's latency assuming the CP runs at 1.5 GHz~\cite{riscv-story} and the CP's
memory access latency is 31 cycles~\cite{kotra2021increasing}.
Given the maximum of 32 software streams, the CP takes $\approx$0.32 $\mu$s to read or write the necessary queues.
Finally, we estimate the predictor overhead by executing it on a CPU with similar specifications to the CP.
Collectively, the total time for the CP to inspect, predict, and write queues is 8 $\mu$s 
(implications discussed in Section~\ref{subsec:eval-over}).
Overall, this setup closely mimics executions on real GPUs, since we add \DESIGN{}'s overheads to runtimes from a real GPU for each given GEMM.

\subsection{Configurations}
\label{subsec:meth-configs}

Since our experiments use a real GPU (Section~\ref{subsec:meth-setup}), we can only perform apples-to-apples quantitative comparisons against other strategies that run on real GPUs (we qualitatively compare against other schemes in Section~\ref{sec:relWork}).
We evaluate the following configurations:
\textbf{sequential} uses the isolated tuning and executes all GEMMs sequentially, \textbf{default} uses isolated tuning and baseline GPU to execute all available GEMM  (via streams) concurrently given GPU resources; \textbf{Globally optimized-Kernels} (\textbf{GO-Kernels}) uses global resource-aware tuning and baseline GPU; \textbf{\DESIGN{}} uses \textbf{GO-Kernels} and \textbf{dynamic logic} at CP to predict the appropriate CD; and \textbf{Oracle} uses GO-Kernels and always chooses the right CD, including sequential execution, if no CD provides $\ge$ 5\% benefit; \textbf{CU-Partition} uses CU masking~\cite{OtternessAnderson2020-cuMask} to statically partition CUs across streams; \textbf{Resource-Partition} statically partitions CUs, LLC, and memory bandwidth across streams~\cite{mig, mxgpu};\footnote{Since our GPU only supports partitioning CUs, we simulate $n$P concurrent GEMMs for \textbf{Resource-Partition} by executing a single GEMM with 1/$n$ CUs, 1/$n$ LLC (by reducing cache size), and 1/$n$ memory bandwidth (by varying memory clock frequency (MCLK)). This model is optimistic, since partitions usually have fewer resources than the overall GPU~\cite{mig}. Furthermore, since our setup can only halve MCLK, we only include 2P results for \textbf{Resource-Partition} and provide optimistic projections for higher CDs (Section~\ref{subsec:eval-mig}).}
We also evaluated Rammer~\cite{MaXie2020-rammer} and ElasticKernels~\cite{pai2013improving}.
However, ElasticKernels does not support kernels that use LDS, which all of our GEMMs do, and our baseline outperformed Rammer by 88\%, which only uses ROCm 3.5.
Thus, we do not show results for Rammer.
Finally, in Section~\ref{subsec:eval-veltair} we evaluate the impact of applying VELTAIR's GEMM optimizations, which were originally designed for CPUs, to GPUs.

\begin{figure*}[tb!]
  \centering
  \includegraphics[scale=0.8]{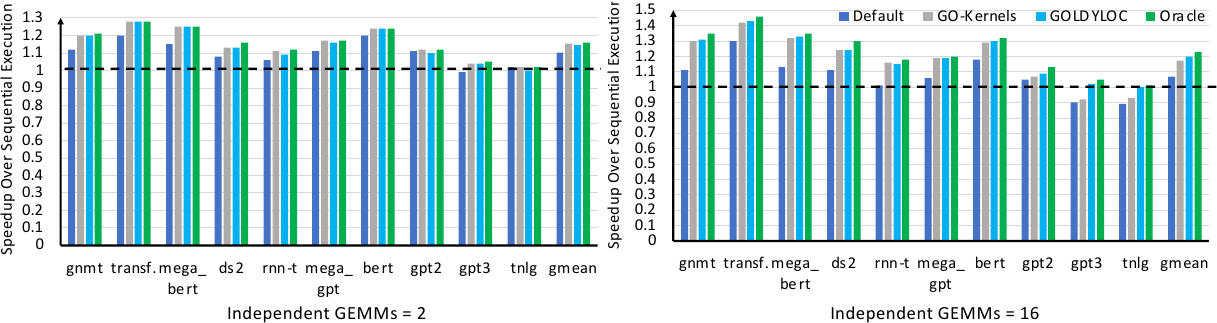}
  \caption{Per-app GEMMs geomean speedups with 2 (left), and 16 (right), independent GEMMs}
  \label{fig:eval}
\end{figure*}

%% file: tables/bmks-new.tex
{\scriptsize
  \begin{tabular}{|c|c|l|l|}
    \hline
    \textbf{Network} & \textbf{Abbreviation} & \textbf{Hyperparameters} & \textbf{Input Params}                                        \\ \hline
    GNMT~\cite{WuSchuster16-seq2seq}                         & gnmt        & H=512;1024          & B=64;128;256;512                    \\ \hline
    DeepSpeech2~\cite{AmodeiAnubhai2016-deepSpeech2}         & ds2         & H=800               & B=64;128;256                        \\ \hline
    RNN-T~\cite{he2018streaming}                             & rnnt        & H=2048              & B=64;128;256;512                    \\ \hline
    Transformer~\cite{VaswaniShazeer17-attention}            & transformer & H=512;1024          & Tokens=512;1024;2048;4096;3072;8192 \\ \hline
    BERT~\cite{DevlinChang18-bert}                           & bert        & H=768;1024          & Tokens=2048;3072;4096;8192          \\ \hline
      GPT-2~\cite{RadfordWu2019-gpt2}                        & gpt2        & H=1280;1600         & Tokens=2048;3072;4096;8192          \\ \hline
      GPT-3~\cite{BrownMann2020-gpt3}                        & gpt3        & H=4096;5140         & Tokens=2048;3072;4096;8192          \\ \hline
      Megatron-LM\_BERT~\cite{ShoeybiPatwary2019-megatronlm} & mega\_bert  & H=1024;2048;2560    & Tokens=2048;3072;4096;8192          \\ \hline
      Megatron-LM\_GPT~\cite{ShoeybiPatwary2019-megatronlm}  & mega\_gpt   & H=1920;3072         & Tokens=2048;3072;4096;8192          \\ \hline
      Turing-NLG~\cite{Microsoft2020-tnlg}                   & tnlg        & H=4256              & Tokens=2048;3072;4096;8192          \\ \hline
  \end{tabular}
}

%% file: 07_evaluation.tex
\section{Results}
\label{sec:eval}

Figure~\ref{fig:eval} shows \DESIGN{}'s benefits over sequential execution for the non-resource partitioned configurations.
Due to space constraints, we only show scenarios with 2 and 16 independent GEMMs (4 and 8 IG's benefits fall in between).
Overall, \DESIGN{}'s geomean benefits increase with more independent GEMMs.
However, the speedups vary considerably for GEMMs across applications.
       
\subsection{Exploiting Concurrency (default)}
\label{subsec:eval-default}

With two independent GEMMs, \textit{default} provides 10\% geomean speedup over executing them sequentially.
However, for almost all GEMMs, further increase in independent GEMMs do not always improve throughput and cause severe slowdowns for GEMMs from large hyperparameter applications (e.g., gpt2, tnlg).
Thus, naively executing all available GEMMs concurrently without tuning for concurrency leads to low speedups.  Moreover, \textit{default}'s geomean speedup across all GEMMs drops (10\% to 7\%) as concurrency increases to 16 independent GEMMs.

\noindent
\textit{\textbf{Result-1:} Naively executing GEMMs concurrently without tuning for concurrency provides small speedups on average.  Moreover, the benefits decrease as concurrency increases.}

\subsection{Globally Optimized (GO)-Kernels}
\label{subsec:eval-lib}

Since \textit{GO-Kernels} are optimized for global resources, they considerably improve performance over \textit{default} (Figure~\ref{fig:eval}) and enable higher CDs than \textit{default}.

\noindent
\textbf{GO-Kernel Properties}: Each GEMM, given its input properties, makes unique trade-offs under resource constraints to pick a uniquely different kernel than its isolated counterpart.
However, there are two key trends: fewer/partial waves and reduced global memory requests. 
In many cases, GO-Kernels have a larger tile size than their isolated counterpart.
Larger tiles improve LDS reuse, reducing LLC/memory requests and thus contention. 
While larger tile size also decreases the total \#WGs, it can increase per-WG resource requirements (e.g., LDS).
Thus, GO-Kernels also change other kernel features to balance performance and per-WG requirements and limit the drop in per-CU occupancy. This combination reduces \#waves and improves overlap. GO-kernels can also have a relatively smaller tile size, 
but also a higher occupancy which also reduces the kernel's \#waves.

\begin{figure}[tb!]
    \centering
    \includegraphics[width=\columnwidth, trim={0cm 0cm 0cm 0cm}]{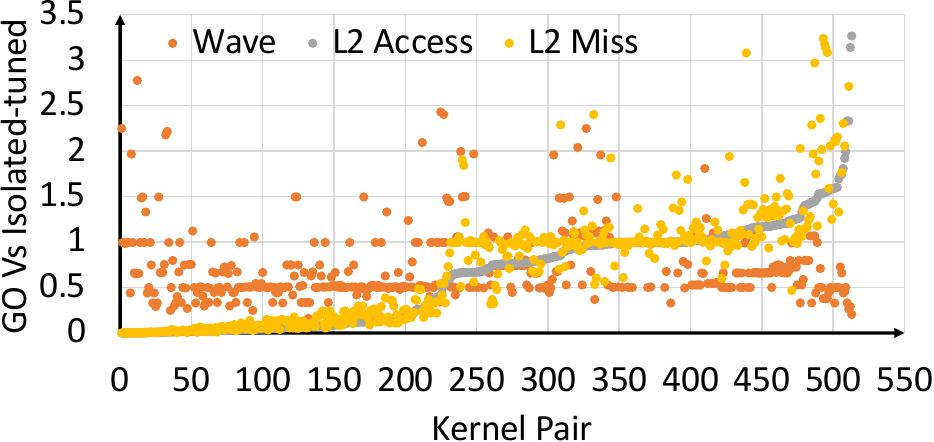}
    \vspace{-2ex}
    \caption{Globally optimized (GO)-Kernel properties.}
    \label{fig:conc_aware_properties}
    \vspace{-2ex}
\end{figure}

Figure~\ref{fig:conc_aware_properties} plots the ratio of \#waves and per-wave LLC accesses/misses in GO-Kernels vs. isolated kernels.
The ratios are largely $<1$, indicating that GO-Kernels have fewer waves and LLC accesses/misses than their isolation-tuned counterparts, making them better for globally sharing resources (Section~\ref{subsec:challs-lib}).
Occasionally (right side of graph), \#waves decrease and LLC activity significantly increases but the latter's absolute values are very small. 
Thus, \DESIGN{}'s resource-constrained tuning properly models concurrent execution environments.

\noindent
\textit{\textbf{Result-2}: Global resource-aware, \textit{GO-Kernels} uniquely differ from their isolated counterparts.}

\noindent
\textit{\textbf{Result-3}: \textit{GO-Kernels} better balance resource requirements, execute in fewer \#waves, and have lower global memory traffic compared to their isolated counterparts.}

\noindent
\textbf{GO-Kernels Benefits}: 
In CD=2P, GO-Kernels have a maximum speedup of 52\% over \textit{default} and provide more than 20\% and 10\% speedup for 11\%, and 24\% of the 410 GEMM sizes, respectively.
Moreover, unlike \textit{default}, GEMM sizes that did not benefit from GO-kernels with 2P do benefit at higher CDs; 53\% of GEMMs in 16P (vs. 34\% in 2P) benefit from GO-kernels.
GO-kernels' benefits over \textit{default} also increase at 16P: 2$\times$ maximum speedup, 25\% of all GEMMs obtain $>$ 20\% speedup, and 43\% of all GEMMs obtain $>$ 10\% speedup.

\begin{figure*}[tb!]
\minipage{0.43\textwidth}
  \includegraphics[width=0.76\linewidth, trim={1cm 16cm 30.5cm 0cm}]{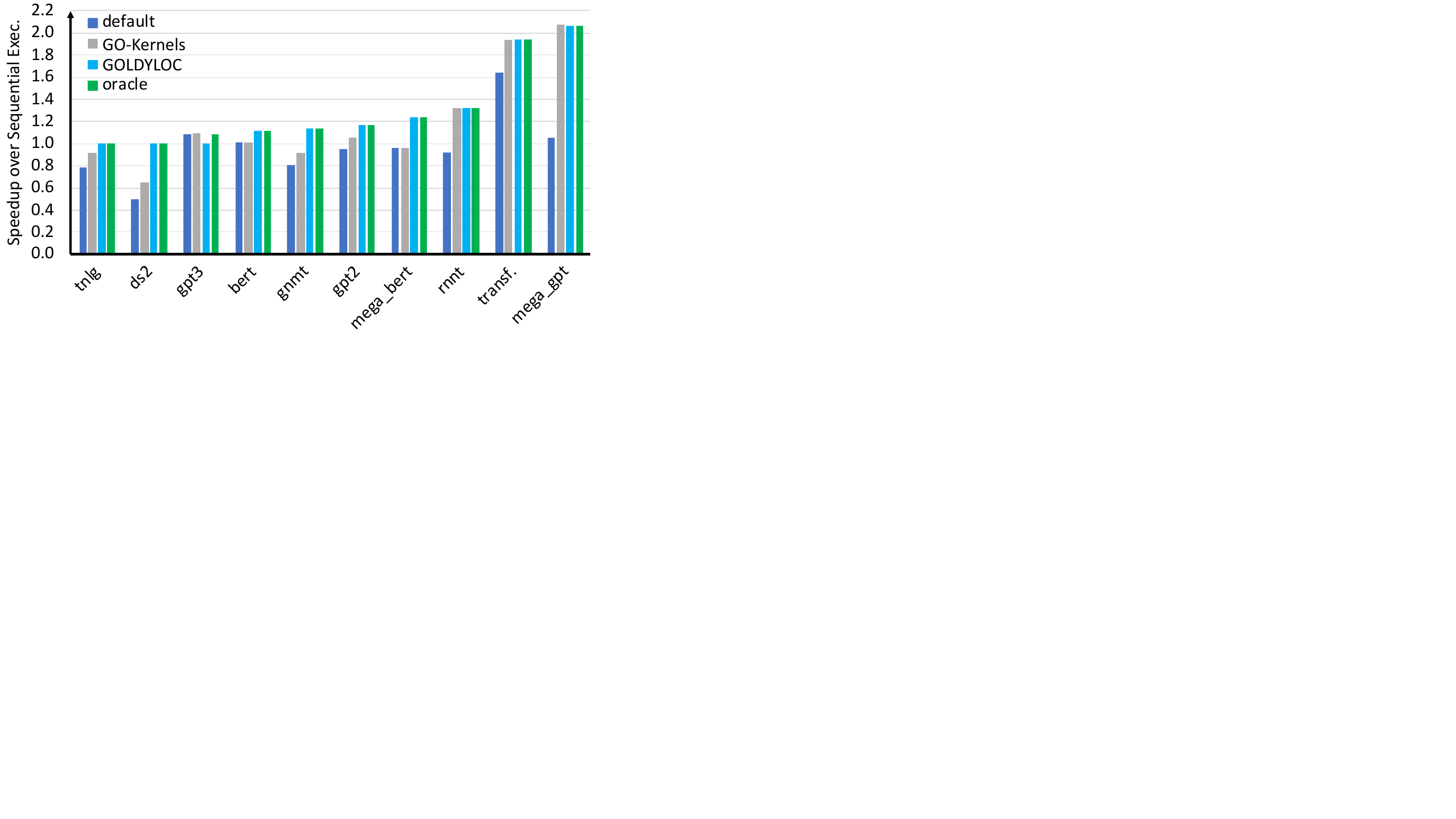}
 \caption{Select GEMMs with CD=16.}\label{fig:individual_gemm_speedup}
\endminipage\hfill
\minipage{0.56\textwidth}%
    \includegraphics[width=\linewidth]{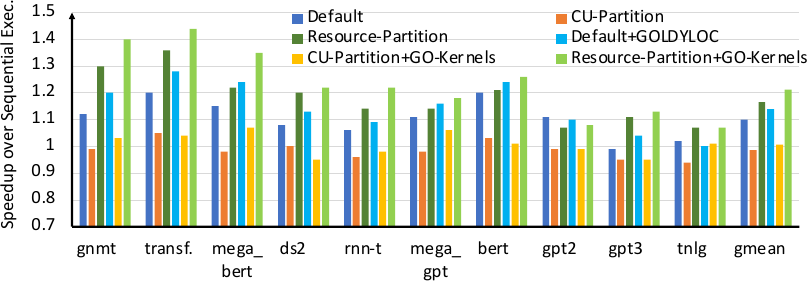}
  \caption{\textit{\DESIGN{}} (CD=2P) with \textit{default} \& CU/Resource partition.} 
  \label{fig:eval_resource_partition}
\endminipage
\end{figure*}

Since not all GEMMs benefit from GO-Kernels (Section~\ref{sec:prop-concur-aware-GEMM-library}), the per-application benefits depend on how many of an application's GEMMs use GO-Kernels and the extent of their benefits (Figure~\ref{fig:eval}).
Most gnmt, transformer, and mega\_bert GEMMs prefer GO-Kernels and achieve higher speedups over \textit{default} and \textit{sequential} execution: 
7-9\% and 20-28\% geomean speedup for 2P and 11-20\% and 30-42\% for 16P.
For applications with few GEMMs that prefer RC-tuning, \textit{GO}'s benefits over \textit{default} are only up to 5\% geomean for CD=2P, but 9-17\% geomean for CD=16P.
Finally, large-dimension GEMMs from large networks (e.g., gpt2, tnlg) are often compute-bound.
RC-tuning for such GEMMs results in kernels whose slowdowns due to limited resources outweigh any benefit from sharing.
Thus, they do not benefit from GO-Kernels and instead require dynamic control (Section~\ref{subsec:eval-concur}).
Across all GEMMs in Figure~\ref{fig:eval}, GO-Kernels achieve 5\% and 10\% geomean speedups over \textit{default} for CDs of 2P and 16P, respectively.
For 4P and 8P CDs, GO-Kernels achieve up to 1.7$\times$ and 2$\times$ speedups, respectively, with 9\% geomean speedups.
Overall, \textit{GO-Kernels}'s benefits are large for small- and medium-sized workloads and increase at higher CDs.
Thus choosing globally optimized kernels is important.

\noindent
\textit{\textbf{Result-4}: \textit{GO-Kernels}' benefits are high for small- and medium-sized workloads, and their benefits increase at higher CDs.}

\subsection{\DESIGN{}}
\label{subsec:eval-concur}

At low levels of concurrency (e.g., 2P), \textit{GO-Kernels} often execute concurrently without heavy contention.
Thus, \textit{\DESIGN{}}, which dynamically controls concurrency (Section~\ref{sec:prop-strategy-predictor}) often provides 
no additional benefits for two independent GEMMs.
However, its benefits increase as available independent GEMMs increase. 
Additionally, large compute-bound GEMMs in gpt2, gpt3, and tnlg suffer at CDs $>2$ because their large per-WG data increase LLC thrashing for more than two concurrent GEMMs.
\textit{\DESIGN{}} accurately predicts this, improving overall performance by 10\% over \textit{GO-Kernels}.
Moreover, \DESIGN{} mispredictions only hurt 7\% of GEMMs (Section~\ref{subsec:eval-accuracy}).
Overall, \DESIGN{} improves performance by up to 35\% (3\% geomean) over \textit{GO-Kernels} and by 5\%, 10\%, 11\%, and 12\% geomean for 2P, 4P, 8P and 16P, respectively, over \textit{default}.

\noindent
\textit{\textbf{Result-5}: \DESIGN{} predicts performant CDs and improves GEMM performance by up to 12\% geomean over \textit{default}.}

\subsection{Range and Distribution of Benefits}
\label{subsec:eval-range}

To demonstrate the range of \textit{\DESIGN{}}'s benefits, Figure~\ref{fig:individual_gemm_speedup} plots their speedups for 16 independent GEMMs for a few GEMM sizes.
In the best cases (rnnt, transformer, mega\_gpt GEMMs), \textit{GO-Kernels} improves performance (up to 2$\times$).
In others (tnlg, ds2, bert, gnmt, gpt2, mega\_bert), \textit{GO-Kernels} provides little benefit, but \textit{\DESIGN{}} selects a more performant CD.
In the worst case (gpt3), \textit{GO-Kernels} do not help, and \textit{\DESIGN{}} mispredicts,  
hurting performance.  
Compared to \textit{default}, across 410 GEMMs \textit{\DESIGN{}} improves 64\% of cases, has no impact on 29\%, and degrades performance only in 7\% of cases.
Thus, \textit{\DESIGN{}} effectively provides benefits across many different GEMMs.

\subsection{CP Overheads}
\label{subsec:eval-over}

To avoid increasing the critical path, CP attempts to perform the prediction, packet setup, and queue prioritization (Section~\ref{sec:meth}) in parallel with prior executing kernels.
Thus, the 8 $\mu$s overhead (Section~\ref{subsec:meth-perf}) is incurred only for the initial kernel and if prior kernels are short ($<8\mu$s).
We study kernel runtime distributions (including non-GEMMs) of several DNNs and all but two kernels have runtimes greater than 8 $\mu$s. 
Thus, the latency can be hidden without impacting end-to-end time.

\noindent
\textit{\textbf{Result-6}: \DESIGN{}'s overheads are small and can be hidden.}

\subsection{Logistic Regression Model Accuracy}
\label{subsec:eval-accuracy}

\DESIGN{}'s logistic regression-based model 
\textit{accuracy} for 2, 4, 8, and 16 available GEMMs is 82\%, 70\%, 62\% and 47\%, respectively. 
Although \textit{\DESIGN{}}'s accuracy decreases for higher number of available GEMMs, which have more output classes, when it is wrong for these scenarios often multiple CDs provide similar (better than \textit{default}) performance.
Thus, it still selects a high-performance CD and provides most of \textit{Oracle}'s benefits (within 3\% geomean).
However, training with a more exhaustive set of GEMMs 
could further improve accuracy and reduce the (small) gap between \textit{\DESIGN{}} and \textit{Oracle}.

\subsection{Heterogeneous GEMMs \& Batched-GEMMs}
\label{subsec:eval-heterog}

Thus far we evaluated \DESIGN{} with homogeneous concurrent GEMMs.
However, \DESIGN{} also improves performance for heterogeneous concurrent GEMMs, where the concurrent GEMMs have unique input sizes.
For brevity we only consider two unique GEMMs, although this is representative of most concurrent backprop GEMMs resulting from independent gradient and error calculations. 
The heterogeneity-agnostic \textit{GO-Kernels} provide 3-10\% geomean speedup over \textit{default} for CD=2-16P.
Extrapolating \textit{\DESIGN{}}'s prediction logic for heterogeneity provides up to 5\% additional speedup for CD=16.
For 16 independent GEMMs, the CP executes all concurrently only if both unique GEMMs prefer 16P.
If not, the CP schedules two sets of 8 independent homogeneous GEMMs.
Overall this provides 15\% geomean speedup over \textit{default} for 16P.

\DESIGN{} also helps with heterogeneous concurrent batched-GEMMs (B-GEMMs)~\cite{bgemm}.
B-GEMMs execute many small, independent, same-sized GEMMs in one kernel~\cite{jhurani2015gemm,abdelfattah2016performance}.
For example, Transformers execute independent B-GEMMs to process variable-length inputs.
Applying \textit{GO-Kernels} to 2P and 4P heterogeneous B-GEMMs provides up to 1.94$\times$ and 1.5$\times$ speedups, and geomean speedups of 5\% and 8\%, respectively, over \textit{default}.

\noindent
\textit{\textbf{Result-7:} \DESIGN{} accelerates heterogeneous concurrent GEMMs by 15\% geomean over default in 16P.}

\noindent
\textit{\textbf{Result-8:} \DESIGN{} accelerates heterogeneous concurrent 
strided batched-GEMMs 
by 8\% geomean over \textit{default} in 4P.}

\subsection{Reduced Precision}
\label{subsec:eval-prec}

\begin{figure}[!tb]
  \centering
  \includegraphics[width=\columnwidth, trim={0cm 0.5cm 0cm 0cm}]{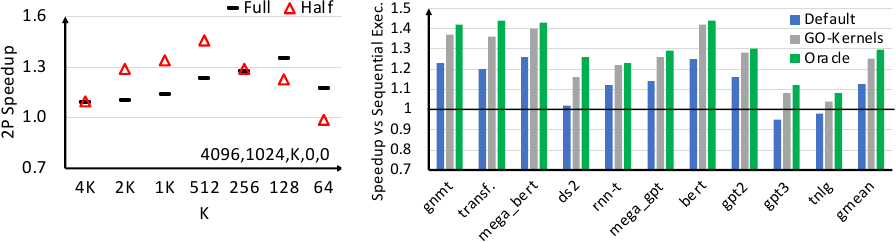}
  \caption{FP16 (a) vs. FP32 2P concurrency with varying GEMM sizes  (b) 16P benefits with \DESIGN{}-Kernels.}
  \label{fig:fp16_data}
\end{figure}

In Figure~\ref{fig:fp16_data}, we evaluate \DESIGN{} with FP16 GEMMs~\cite{ micikevicius2018mixed, wang2018training, amd-cdna-arch, NV-DL-Perf, pytorch-amp, ott2018scaling,FowersOvtcharov2018-brainwave}.
Since FP16 throughput on the same device is usually higher than FP32's, its peak concurrency speedup also increases (Figure~\ref{fig:fp16_data}(a)).
The curve in Figure~\ref{fig:fp16_data}(a) also shifts left with FP16, 
implying more potential benefits with larger sizes.
While concurrency benefits with larger (e.g., tnlg) GEMMs could be higher in FP16 than FP32, it is not observed due to isolated tuning. As shown in Figure~\ref{fig:fp16_data}(b), \textit{GO-Kernels} speeds up 16P GEMMs with gpt2, gpt3, and tnlg sizes by 10\%, 14\%, and 6\% geomean, respectively.

\noindent
\textit{\textbf{Result-9:} \DESIGN{} benefits increase for large GEMMs at reduced precision.}

\subsection{\DESIGN{} with Resource Partitioning}
\label{subsec:eval-mig}

We also evaluate \textit{GO-Kernels}'s impact on resource partitioned configurations in Figure~\ref{fig:eval_resource_partition}.
\textit{CU-Partition} is often worse than \textit{default} due to memory resource contention and underutilized CU resources (partial wave within a partition).
Conversely, the optimistic \textit{Resource-Partition}'s dedicated memory resources help it outperform \textit{default} (similar to
prior work~\cite{tan2021gpupool,mig-eval}).
Nevertheless, partitioning resources defines constraints, making global resource-aware optimizations even more important.
Thus, as shown in Figure~\ref{fig:eval_resource_partition}, for CD=2P,  re-using \textit{GO-kernels} tuned for \textit{default} provides up to 1.4$\times$ and 1.6$\times$ (3\% and 4\% geomean) benefits over \textit{CU-Partition} and \textit{Resource-Partition} for CD=2P, respectively. \textit{GO-kernels} when tuned for these configurations increase these benefits to 6\% and 9\% geomean, respectively. Finally, speedups increase to 5-22\% for 4-8P over \textit{CU-Partition} and 7\% over an optimistic 4P \textit{Resource-Partition}.

\noindent
\textit{\textbf{Result-10:} Partitioning resources improves performance versus \textit{default} but still benefits from globally optimized kernels.}

\subsection{End-to-end Speedups}
\label{subsec:eval-end-to-end}

RNNs and Transformers have significant intra-network parallelism.
For example, GNMT (H=1024) can execute up to eight (layer) GEMMs in parallel.
Thus, \textit{\DESIGN{}} speeds up its iterations by 14\% and 13\% (for batch size 128 and 256, respectively) over \textit{default}.
\DESIGN{} also speeds up parallel Attention B-GEMMs and gradient GEMMs in Transformers: \DESIGN{} speeds up BERT's iteration times by 5-12\% over \textit{default}.

\subsection{GEMM Fusion}
\label{subsec:eval-fusion}

Although GEMM fusion~\cite{Appleyard2016-optRNNGPU, FilipovivcMadzin2015-blasCUDAFusion, LiZheng2020-horizFus, cudnn, SivathanuChugh2019-astra}
improves throughput, it is only applicable if GEMMs share inputs or the application sums all the GEMMs' outputs.
Its benefits also saturate as matrix sizes grow.
For example, in Transformers the input projection for QKV GEMMs can be fused.
However, fusion's benefits decrease as the input activation size (determined by batch-size and sequence-length) increases~\cite{PatiAga2022-demystifying}.
Furthermore, concurrency with \textit{\DESIGN{}} can often outperform fusion.
For instance, in QKV layer of BERT, concurrently executing two GEMMs of this layer (in both forward and backward prop) with \DESIGN{} achieves 7\% better speedups than fusing them. This is likely due to GO-Kernels' fewer memory accesses, fewer \#waves and/or fewer total instructions as compared to the fused kernels.
In RNNs, fusion also determines available parallelism amongst other operations.
Although fully fusing all possible GNMT GEMMs (Section~\ref{subsec:eval-end-to-end}) improves performance by 19\% over \textit{sequential}, it serializes other, smaller GEMMs (Section~\ref{sec:char}), causing benefits to saturate beyond fusing eight GEMMs.
Thus, \textit{\DESIGN{}} outperforms fusion by 10\%. 
These results highlight how dynamic selection of fusion versus concurrency for potentially fusable GEMMs can further improve performance of independent GEMMs.

\noindent
\textit{\textbf{Result-11:} \DESIGN{} speeds up GNMT by 14\% over \textit{default}, and by 10\% over maximum GEMM fusion.}

\subsection{Comparing \DESIGN{} to VELTAIR}
\label{subsec:eval-veltair}

Prior work like VELTAIR~\cite{liu2022veltair} exploits concurrency on CPUs.
However, while concurrent executions on different substrates (CPUs, GPUs) consider similar factors (e.g., parallelism, reuse), the trade-offs and outcomes often differ.
GPU CUs have large register files and dedicated specialized memories (e.g., LDS), while CPU cores have large shared caches.
Such differences can lead to different outcomes when selecting appropriate concurrent implementations. VELTAIR prefers smaller tiles because maximizing reuse via larger tiles increases shared LLC contention and causes poor concurrent performance in CPUs.
Conversely, GPUs prefer larger tiles as it improves LDS reuse and reduces memory traffic -- improving concurrent performance. Consequently, when we applied VELTAIR's principles to GPUs, we found its smaller tiles hurt concurrent GEMM performance by 17-26\% geomean for CDs of 2-16 compared to \textit{\DESIGN{}}'s larger tiles. Thus, VELTAIR does not always select high performing GPU GEMMs.

%% file: 08_discussion.tex
\section{Discussion}
\label{sec:disc}

\subsection{Non-GEMM GPU Kernels}
\label{subsec:disc-nonGEMM}

DNNs also have interspersed non-GEMM operations, including element-wise adds, multiplies, reductions, and activation functions. Most of these operations are bottlenecked by memory accesses. Accordingly, software frequently uses optimizations such as kernel fusion to fuse series of such operations into a single kernel, often with preceding GEMMs, to avoid redundant global memory accesses. This significantly reduces the runtime of non-GEMM operations. Thus, as mentioned in Section~\ref{sec:meth}, we focus on GEMMs because they constitute the majority of runtime in DNNs. Furthermore, unlike non-GEMM kernels, libraries are rigorously tuned for different GEMM input sizes, leaving significant room for improvement in case of concurrent execution.

We also evaluated \DESIGN{} with a GEMM concurrently executing with a non-GEMM (2P). We execute non-GEMMs (element-wise adds) with input sizes that match the concurrent GEMM's output as non-GEMMs in DNNs usually operate on GEMMs' output or activations. On average, GEMMs with GO-Kernels speed up the concurrent GEMM-non-GEMM executions by only 3\% over \textit{default}. However, cases with memory-bound GEMMs (GEMMs with small $N$ and $K$ dimensions) have much larger (average 10\%) benefits.
Finally, \DESIGN{}, by restricting concurrency, provides larger benefits (8\% geomean) in GEMM-non-GEMM cases, especially for those with memory-bound GEMMs (33\%).
Thus, \DESIGN{} also helps concurrent execution of GEMMs with other non-GEMM GPU kernels.

\subsection{Sparsity}
\label{subsec:disc-sparse}
Prior work has shown that significant sparsity exists in many of these networks~\cite{correia2019adaptively, Narang2017-sparseRNNs, Zhu2018-sparsePRNN,Parashar17-scnn, HanKang2017-ese}.
Leveraging sparsity is especially useful for very large networks with large parameter matrices.
Although evaluating the additional behavior when exploiting sparsity is beyond the scope of this paper, we expect concurrency will become more important as sparsity reduces the amount of computation in GEMMs.

\subsection{Additional Resource Constraints \& Overheads}
\label{subsec:disc-rcs}

For tuning, we evaluate only two additional (RC) configurations (GPU/2 and GPU/4 in Section~\ref{sec:prop-resource-constrained-tuning}).
Adding GPU/4 to GPU and GPU/2 improved performance for 34\% of GEMMs.
Stricter RCs (GPU/8 and GPU/16) provided little benefit, likely because kernels become prohibitively slow at such low resources, limiting concurrency benefits.
However, given rapid rate GPU compute is scaling, stricter RCs may become necessary.
We also tried constraining memory bandwidth (BW) using memory clock frequency (MCLK) as a proxy (constraining BW via specific memory allocations was beyond the scope of the paper) but found limited additional benefits.
We believe this is because constraining MCLK also impacts memory latency, which may not be representative of concurrent execution environments.
Constraining additional shared resources may provide more concurrency-amenable kernels.
Finally, not all GEMM sizes require kernels tuned for all three configurations studied.
Some only prefer GPU/2 and some do not prefer RC configurations altogether.

\begin{figure}[bt!]
  \centering
  \includegraphics[width=0.8\columnwidth]{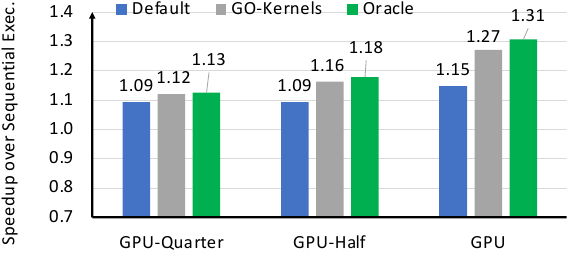}
  \caption{CD=4P speedups for multiple GPU configurations.}
  \label{fig:factors_config}
\end{figure}

\subsection{Scaling GPUs Configuration}
\label{subsec:disc-config}

Since GPUs are rapidly scaling, we study \DESIGN{}'s benefits by changing hardware resources.
Specifically, we compare \DESIGN{} on \textit{GPU-Quarter} (32 CUs, 2 MB LLC), \textit{GPU-Half} (64 CUs, 4 MB LLC), and the original \textit{GPU} (120 CUs, 8 MB LLC). Figure~\ref{fig:factors_config} shows that \DESIGN{}'s benefits are higher as GPUs scale up: benefits increase from 3\% in \textit{GPU-Quarter} to 12\% in \textit{GPU}.
Scaling GPU compute with fixed memory bandwidth also increases contention, making \DESIGN{} more effective.

\subsection{Reducing Tuning Overhead}
\label{sec:prop-taming-tuning-overhead}

Although GO-Kernel's overhead is a one-time cost, predicting a GEMM's preferred RC configuration (PRC) for a given CD can reduce this overhead.
We examined K-Nearest Neighbor (KNN)-based classification to predict a new GEMM's PRC based on the PRC of K closest GEMMs by Euclidean distance.
We exhaustively tune for 20\% of GEMMs (Section~\ref{sec:prop-tuning} and predict the PRC for the remaining 80\%, using size ($M * N$) and \textit{default} kernels' tile size to determine closeness.
Along with dynamic control, it still improves performance over \textit{default} by 2-9\% overall (for CD$=$2-16P).

\subsection{Other DNNs}
\label{subsec:disc-dnns}

\DESIGN{} also helps 
CNNs, Multilayer Perceptron (MLP) layers in recommendation models~\cite{GuptaHsia2020-deepRecSys}, and Graph Neural Network's~\cite{sun2022cognn}.
Their inter-GEMM parallelism arises from gradient descent, checkpointing, and multi-instance runs (Section~\ref{sec:char}).
For example, \DESIGN{} speeds up MLPerf's ResNet-50 and DLRM independent GEMMs 
by up to 21\% and 36\%, respectively. Additionally, Mixture-of-Expert models also increase scope for concurrent executions by activating multiple layers (experts) concurrently, each operating on a subset of input data~\cite{HwangWei2024-moe} and can benefit from \DESIGN{}.

%% file: 09_related.tex
\section{Related Work}
\label{sec:relWork}

Table~\ref{tab:related} compares \DESIGN{} to prior work 
and shows that \DESIGN{} is the only approach to provide all four important features.
Moreover, to the best of our knowledge, no prior work leverages the CP to improve concurrency.

\noindent
\textbf{Other Devices}:
Concurrency helps 
maximize device resources.
Similar to \DESIGN{} which improves GPU concurrency, VELTAIR~\cite{liu2022veltair} optimizes multi-tenancy on CPUs, while MAGMA~\cite{kao2022magma} and HERALD~\cite{kwon2019herald} focus on accelerators.
Although these prior works have a similar goal, their optimizations differ since they target latency-oriented CPUs~\cite{liu2022veltair} or dataflow-based accelerators~\cite{kao2022magma,kwon2019herald}.
Moreover, these architectural differences often result in different designs (Section~\ref{subsec:eval-veltair}).

\noindent
\textbf{GPU Scheduling}:
Other works improve GPU concurrency via better wavefront~\cite{cruise, Rogers2012, cawa,dwf, gpu-trace-schedule, gputhread, RogersOConnor2013-daws,stall-aware,simd_sched_gra, owl, micro_2_lvl,pats,LiuYang2015-saws} and queue~\cite{ChenYang2017-prophet, ChenYang2016-baymax, GaoYu2018-batchmaker, HolmesMawhirter2019-grnn,gpusync, KatoLakshmanan2011-timeGraph, YehSinclair2021-lax,adriaens2012case} scheduling.
Thus, they \textit{dynamically} manage intra- and/or inter-process concurrency.
However, unlike \DESIGN{}, these approaches only consider isolated, globally suboptimal kernels. Additionally, \DESIGN{} could also be integrated with wavefront scheduling optimizations.

\begin{table}[tb!]
\centering
\setlength\tabcolsep{2pt}
\resizebox{\columnwidth}{!}{%
\input{tables/related_table}
}
\caption{Comparing \DESIGN{} to prior work.}
\label{tab:related}
\vspace{-6ex}
\end{table}

\noindent
\textbf{Globally optimized kernels}:
Prior work also designed GPU kernel implementations for concurrency.
For example, Rammer~\cite{MaXie2020-rammer} and ElasticKernels partially design globally-optimized kernels.
As discussed in Section~\ref{sec:intro}, former does not support key kernel features from BLAS libraries while latter does not support LDS-heavy GEMMs.
Moreover, in Section~\ref{subsec:meth-configs} we show that our baseline outperforms Rammer.
Batched-GEMMs~\cite{bgemm,jhurani2015gemm,abdelfattah2016performance}
execute small independent GEMMs within a kernel but require expensive data layout/application changes and are not applicable to heterogeneous and inter-application GEMMs.
Additionally, Section~\ref{subsec:eval-heterog} shows that \DESIGN{} helps concurrent batched-GEMMs.
Finally, unlike \DESIGN{}, none dynamically control concurrency.

%% file: tables/related_table.tex
{\scriptsize
\begin{tabular}{|c|c|c|c|c|}
\hline
\textbf{Approach / Features}             & \textbf{GPU Support} & \textbf{Globally Optimized} & \textbf{Dynamic Control} & \textbf{No App. Changes}   \\ \hline

\textbf{Herald~\cite{kwon2019herald}}        & \textcolor{red}{X}                      & \textcolor{red}{X}                                                                                    & \textcolor[HTML]{00A99A}{\checkmark}                                                                                     &  \textcolor[HTML]{00A99A}{\checkmark}                                                                                               \\ \hline
\textbf{Magma~\cite{kao2022magma}}        & \textcolor{red}{X}                      & \textcolor{red}{X}                                                                                    & \textcolor[HTML]{00A99A}{\checkmark}                                                                                     &  \textcolor[HTML]{00A99A}{\checkmark}                                                                             \\ \hline
\textbf{VELTAIR~\cite{liu2022veltair}}               & \textcolor{red}{X}                    &         \textcolor[HTML]{00A99A}{\checkmark}                                & \textcolor[HTML]{00A99A}{\checkmark}                                                                                          & \textcolor[HTML]{00A99A}{\checkmark}                                                            \\ \hline
\textbf{Queue Schedulers}     & \textcolor[HTML]{00A99A}{\checkmark}                   & \textcolor{red}{X}                                                                                      & \textcolor[HTML]{00A99A}{\checkmark}                                                                                           & \textcolor[HTML]{00A99A}{\checkmark}                                                                                                                                        \\ \hline
\textbf{Wavefront Schedulers} & \textcolor[HTML]{00A99A}{\checkmark}                   & \textcolor{red}{X}                                                                                      & \textcolor{red}{X}                                                                                              & \textcolor[HTML]{00A99A}{\checkmark}                 \\ \hline

\textbf{Rammer~\cite{MaXie2020-rammer}}               & \textcolor[HTML]{00A99A}{\checkmark}                   & Partial                                                                              & \textcolor{red}{X}                                                                                          & Partial                                                                 \\ \hline
\textbf{Elastic Kernels~\cite{pai2013improving}}      & \textcolor[HTML]{00A99A}{\checkmark}                   & Partial                                                                                & \textcolor{red}{X}                                                                                             &  \textcolor[HTML]{00A99A}{\checkmark}                                                                                                                                       \\ \hline
\textbf{Batched-GEMM~\cite{bgemm}}         & \textcolor[HTML]{00A99A}{\checkmark}                   & Partial                                                                        & \textcolor{red}{X}                                                                                 & \textcolor{red}{X}                                                                                              \\ \hline
\textbf{\begin{tabular}[c]{@{}c@{}}\DESIGN{}\end{tabular}}           & \textcolor[HTML]{00A99A}{\checkmark}                   & \textcolor[HTML]{00A99A}{\checkmark}                                                                                   & \textcolor[HTML]{00A99A}{\checkmark}                                                                                          & \textcolor[HTML]{00A99A}{\checkmark}                                                                \\ \hline
\end{tabular}
}

%% file: 10_conc.tex
\section{Conclusion}
\label{sec:conc}

Applications such as DNN training and inference have abundant opportunities to execute GEMMs concurrently.
Unfortunately, exploiting this concurrency is difficult in GPUs as they use kernels tuned in \textit{isolation}, manage concurrency \textit{statically}, or both.
\DESIGN{} solves this for key GEMM operations by (1) tuning kernels for globally shared resources during concurrency, and (2) extending the GPU's CP to dynamically control how many GEMMs to execute concurrently.
\DESIGN{} improves performance by 2.5$\times$ max (43\% geomean per-app) over sequential execution and 2$\times$ max (18\% geomean per-app) over concurrent execution in current GPUs. Overall, our work demonstrates how co-designing applications, hardware, and the runtime between them can significantly improve efficiency.